\documentclass[letterpaper]{JHEP3}
\pdfoutput=1

\usepackage{graphicx}
\usepackage{caption}
\usepackage{subcaption}
\usepackage{amsmath}
\usepackage{amssymb}
\usepackage{amsfonts}
\usepackage{lscape, graphicx}
\usepackage{cite}       % collapse sequential citations to a range
\usepackage{bm}         % bold math
\usepackage{verbatim}
\usepackage{subfig}

\usepackage{amsmath}
\usepackage{amssymb}
\usepackage{amsfonts}
\usepackage{lscape, graphicx}
\usepackage{cite}       % collapse sequential citations to a range

\usepackage[all]{xy}
\advance\parskip 1.0pt plus 1.0pt minus 2.0pt   % needs some stretchability
\def\href#1#2{#2}	% Avoid  JHEP bug with footnotes in author list

\def\coeff#1#2{{\textstyle {\frac {#1}{#2}}}}
\def\half{\coeff 12}

\def\R{{\mathbb R}}
\def\S{{\mathbb S}}

\def\Z{{\mathbb Z}}

\def\Dslash{{\rlap{\raise 1pt \hbox{$\>/$}}D}}

\newcommand{\beq}{\begin{equation}}
\newcommand{\eeq}{\end{equation}}
\newcommand{\beqa}{\begin{eqnarray}}
\newcommand{\eeqa}{\end{eqnarray}}

\def\ltap{\ \raise.3ex\hbox{$<$\kern-.75em\lower1ex\hbox{$\sim$}}\ }
\def\gtap{\ \raise.3ex\hbox{$>$\kern-.75em\lower1ex\hbox{$\sim$}}\ }
\def\gl{\ \raise.5ex\hbox{$>$}\kern-.8em\lower.5ex\hbox{$<$}\ }
\def\roughly#1{\raise.3ex\hbox{$#1$\kern-.75em\lower1ex\hbox{$\sim$}}}

\title{The $\mathbf{SU(3)}$/$\mathbf{\Z_3}$ QCD(adj)  deconfinement  transition\\ via  the gauge theory/``affine" XY-model duality}

  \author
{
    {
    Mohamed M. Anber,\footnote{\email{manber@physics.utoronto.ca}} ~
   Scott Collier,\footnote{\email{scott.collier@utoronto.ca}}  ~and   Erich Poppitz\footnote{\email{poppitz@physics.utoronto.ca}}
           \\ 
           {Department of Physics, University of Toronto,
    Toronto, ON M5S 1A7, Canada}
           
            }
    }%
    
    \abstract{ 
    \smallskip
    
Earlier, two of us  and M.~\" Unsal   \cite{Anber:2011gn} showed that  a class of 4d  gauge theories, when compactified on a small spatial circle of size $L$ and considered at temperatures $\beta^{-1}$ near the deconfinement transition, are dual to 2d ``affine'' XY-spin models. We exploit this duality to study the deconfinement phase transition in  $SU(3)$/$\Z_3$ gauge theories with $n_f > 1$ massless adjoint Weyl fermions, QCD(adj) on $\R^2 \times \S^1_\beta \times \S^1_L$.
The dual ``affine" XY-model describes  two 
``spins"---compact scalars taking values in the $SU(3)$ root lattice. The  spins  couple  via nearest-neighbor interactions  and are subject to an ``external field"  perturbation preserving  the  topological $\Z^t_3$ and  a discrete $\Z^{d \chi}_3$ subgroup of the anomaly-free chiral symmetry of the 4d   gauge theory. The equivalent Coulomb gas representation of the theory exhibits  electric-magnetic duality, which is also a high-/low-temperature duality. A renormalization group analysis suggests---but is not convincing, due to the onset of strong coupling---that the self-dual point  is a fixed point,  implying a continuous deconfinement transition. Here, we study the nature of the transition via
 Monte Carlo simulations. The  $\Z^t_3 \times \Z^{d\chi}_3$ order parameter, its susceptibility, the  vortex density,  the energy per spin, and the specific heat are measured over a range of volumes, temperatures, and ``external field" strengths (in the gauge theory, these correspond to magnetic bion fugacities). The finite-size scaling of the susceptibility and specific heat we find is characteristic of a first-order transition. Furthermore, for sufficiently large but   still smaller than unity  bion fugacity (as can be achieved  upon an   increase of the $\S_L^1$ size), at the critical temperature we find two distinct peaks of the energy probability distribution, indicative of a first-order  transition, as has been seen in earlier simulations of the full 4d QCD(adj) theory. We end with discussions of the global phase diagram in the $\beta$-$L$ plane for different  numbers of flavors.

    \smallskip
    
    \smallskip
    
       {\small{
     }

}}

\begin{document}

\maketitle

%%%%%%%%%%%%%%%%%%%%%%%%%%%%%%%%%%%
\section{Introduction}

\subsection{Motivation: compactification on $\mathbf{\S^1_L}$  as a theoretical laboratory}

The observation that the dynamics of certain four-dimensional (4d) gauge theories considerably simplifies
 upon compactification on a spatial circle of small size $L$---i.e., upon considering the theories on $\R^{1,2} \times \S^1_L$ instead of $\R^{1,3}$---originated in studies of 4d supersymmetric theories, compactified on $\S^1_L$ from the late 1990's. Many results, some of which were new, on supersymmetric gauge dynamics were obtained by  small-$L$ studies, combined with the ``power of holomorphy" \cite{Seiberg:1996nz}.
In 2007, \" Unsal realized \cite{Unsal:2007vu} that this simplification  transcends supersymmetry and also occurs in a large class of nonsupersymmetric 4d gauge theories, notably the ones with adjoint massless fermions,   QCD(adj). At small $L$, these theories dynamically abelianize and many nonperturbative properties, usually  not amenable to an analytic treatment---confinement, the generation of mass gap, and discrete or abelian  chiral symmetry breaking---become tractable within a theoretically controlled semiclassical approximation.   It was found  in \cite{Unsal:2007vu} that in QCD(adj), the nonperturbative mass gap  is due to a new kind of topological ``molecules"---the ``magnetic bions", carrying zero topological but nonzero magnetic charge. The magnetic bions  populate the vacuum and 
 cause 
confinement of electric charges by  the 3d Debye screening mechanism of Polyakov \cite{Polyakov:1976fu}. It is important to stress that while the long-distance confining dynamics is, indeed, three-dimensional, the mere existence of the magnetic bions is due to the locally 4d nature of the theory (in the 3d $L = 0$ limit there are no magnetic bions). 
These small-$L$ studies showed that a piece of conventional wisdom, that ``confinement is a pure-glue phenomenon,"  does not hold universally---the nature of the topological excitations causing the mass gap crucially depends on the massless fermion content. This dependence was later found to be a generic phenomenon, occurring in a variety of nonsupersymmetric gauge theories with massless fermions, both chiral and vectorlike \cite{Shifman:2008cx,Poppitz:2009uq}.

After the small-$L$ physics of confinement is semiclassically understood, it is natural to also study the thermal behavior of the theory. Not surprisingly, one finds that it also has interesting and novel features. It is well-known that 4d gauge theories undergo a confinement-deconfinement phase transition at some critical temperature. This transition persists upon compactification to small-$L$ (the theories are now considered on $\R^2 \times \S^1_\beta \times \S^1_L$, where $\beta$ is the inverse temperature). The small-$L$  finite-$T$ dynamics  can also be described semiclassically: near the deconfinement transition, the partition function of the gauge theory is that of a 2d Coulomb gas of electrically and magnetically charged particles, interacting via long-range Coulomb (and dual-Coulomb) interactions and by Aharonov-Bohm phase interactions. This is similar to the case of the 3d Polyakov model or deformed pure Yang-Mills theory \cite{Dunne:2000vp}, but because of the ``molecular" nature of the topological excitations causing confinement, the dynamics has a different, rather rich, and yet unexplored  structure. 

In what follows, we will describe the map between the physics of thermal 4d QCD(adj) on $\S_L^1$ and the spin model that we study in this paper (or the equivalent\footnote{The duality between 2d electric-magnetic Coulomb gases and   theories of coupled XY-spins has a long history in the condensed matter literature, see \cite{Jose:1977gm}. However, the ``affine" XY-models dual to QCD(adj), see eq.~(\ref{z3sqrdmodel}), especially the ones for $N_c >3$, are new.}  Coulomb gas). We will not give a derivation, for which we refer the reader to \cite{Anber:2011gn}, but provide this discussion in order to elucidate the relevant scales and the relation of the quantities of interest in QCD(adj) to   spin-model observables. We emphasize that the dual spin model is more than  a Landau-Ginsburg effective theory of an order parameter: owing to the small-$L$ calculability, we have  a detailed understanding of how the microscopic physics of the gauge theory is reflected in the affine spin model.
We will also outline the qualitative picture of the deconfinement transition as occurring due to a competition between electric and magnetic degrees of freedom. 
The goal of this paper is to study this picture in more quantitative detail, with emphasis on finding the order of the deconfinement transition by numerical simulation of the dual spin model.

\subsection{Digression: on the small-$\mathbf{L}$/large-$\mathbf{L}$ relation}
\label{digression}

Before continuing with the study outlined above, let us comment on the relation between small- and large-$L$ physics. The magnetic bion mechanism of confinement in QCD(adj) is operative at small but nonzero $L \ll \Lambda_{QCD}^{-1}$,  where it is under full theoretical control. It gives us a remarkable  window of calculability permitting the analytical study of difficult nonperturbative phenomena in locally 4d gauge theories---an opportunity, which (we feel) is interesting to explore on its own.

As $L$ increases at fixed $\Lambda$,  strong coupling sets in, making it difficult (in theories without supersymmetry) to continue to the physically interesting case of large $L$ while retaining theoretical control.\footnote{An exceptional  nonsupersymmetric example is provided by theories which at $L\rightarrow \infty$ run into a Banks-Zaks-like weakly coupled fixed point---as the $n_f=5$ QCD(adj) is believed to---see \cite{Poppitz:2009uq} and Section \ref{phasediagram}.}
In   Seiberg-Witten theory, however, a continuous connection to the 4d mechanism of confinement by monopole condensation exhibited by softly-broken ${\cal{N}}=2$  supersymmetric theories can be made \cite{Poppitz:2011wy}. Thus, an explicit  relation exists between  the two known theoretically controlled mechanisms  of confinement  in  locally   4d nonabelian gauge theories (admittedly both mechanisms are essentially abelian). While there should be   a connection to the 4d mechanisms of nonabelian confinement and chiral symmetry breaking (also nonabelian) in non-supersymmetric theories as well, we feel that it is not likely to be found  by purely analytical means (we base this expectation on  the  complex relation  between the two mechanisms  found in  Seiberg-Witten theory and the crucial role played by supersymmetry in establishing it). 
We finally mention two related recent observations. First, the use of resurgence and trans-series in quantum field theory   \cite{Dunne:2012zk}  may eventually offer a more explicit connection between small- and large-$L$ physics, at least  in particular theories.
Second,   it was recently realized that the novel topological ``molecules" found at small-$L$ (e.g., the already-mentioned ``magnetic bions", the ``center-stabilizing bions", and various kinds of monopole-instantons) may be relevant to 4d physics in another guise, namely play a role in the microscopic description of the thermal deconfinement transition in nonsupersymmetric 4d pure Yang-Mills theory     \cite{Poppitz:2012sw}.

\subsection{Mapping a 4d gauge theory on $\mathbf{\S_L^1}$ to an ``affine" XY model: the melting of 2d-crystals  and $\mathbf{SU(3)}$ QCD(adj)}
\label{su3intro}

In this paper, our   interest  is 
   $SU(3)/\Z_3$ QCD(adj) with $n_f$ massless adjoint Weyl fermions
 on $\R^{1,2}  \times \S^1_L$.
 As shown in \cite{Anber:2011gn}, the theory near the deconfinement transition is described by an ``affine" XY-spin model. The reason this is possible is that the partition functions of both  the spin model and thermal QCD(adj) at small-$L$, for a range of temperatures near the deconfinement transition, can be given a 2d  Coulomb gas representation. The parameters of the two Coulomb gases can then be mapped to each other.  In the rest of this Section, we first briefly review the continuum dynamics of QCD(adj) and then present the dual spin model description.

\subsubsection{Review of the continuum dynamics of QCD(adj) on $\mathbf{\R^2 \times S^1_L \times S^1_\beta}$}
\label{review}

We begin with the zero-temperature case. When the theory\footnote{In this Section, we   will not distinguish between $SU(3)$ and $SU(3)/\Z_3$ gauge theory. The distinction will become relevant  in Section \ref{natural}  where we consider the dual spin model, which has the symmetries of the $SU(3)/\Z_3$ theory. In particular, there is no center symmetry, but a topological $\Z_3^t$ symmetry (also called ``dual center symmetry" \cite{Argyres:2012ka}), see the discussion in  Section \ref{natural}.} is considered at small $L$, $SU(3)$ QCD(adj) dynamically abelianizes. In other words, $SU(3)$ breaks to $U(1)^2$ at a scale $\sim {1\over L} \gg \Lambda_{QCD}$. Thus, the only perturbative excitations relevant to the dynamics at length scales $\gg L$ are   
 the two massless photons in the unbroken Cartan subalgebra of $SU(3)$. These can be dualized to compact scalars (recall that a photon in 3d is dual to a scalar). Some fermions also remain 
massless (the components of the adjoint fermions  along the Cartan generators). 

Nonperturbatively, however, the dynamics is quite complex and various topological excitations, the magnetic monopole-instantons \cite{Polyakov:1976fu} and twisted \cite{Lee:1997vp} monopole-instantons, play a central  role. The small-$L$ nonperturbative dynamics can be described semiclassically because the two $U(1)$ couplings are free and the theory is weakly coupled. The upshot of the studies of \cite{Unsal:2007vu} is that ``molecular" (i.e., correlated) instanton events, the magnetic bions, proliferate in the vacuum and cause the appearance of mass gap and the confinement of electric charges (see also \cite{Anber:2011de, Argyres:2012ka} for recent detailed studies). The nonperturbative mass gap is, up to pre-exponential factors:
\begin{equation}
\label{msigma}
m_\sigma \sim {1\over L}\; e^{ - {4 \pi^2 \over g_4^2(L)}}~. 
\end{equation}
Here, $L$ is the size of $\S^1_L$ and $g_4^2(L)$---the 4d gauge coupling, frozen at the scale $L$. The partition function of the zero-temperature theory can be described as a dilute 3d Coulomb gas of magnetic bions, interacting via mutual ${1\over r}$  ``magnetostatic" potentials.

Next, we consider the theory at finite temperature, specifically in the   regime:
\begin{equation}
\label{temperaturerange}
{1\over L} e^{ - {8 \pi^2 \over 3 g_4(L)^2}} \ll T \ll {g_4^2(L) n_F \over \pi L}~.
\end{equation}
The physical meaning of the upper bound on $T$ in (\ref{temperaturerange}) is that the temperature has to be smaller than the inverse size of the bions, so that bions can be treated as pointlike and non-dissociated objects, while the lower bound implies that the temperature is larger than   the  inverse of the typical distance between bions, so that the bions are described as an effectively 2d gas  \cite{Anber:2011gn}. It turns out that (\ref{temperaturerange}) is the interesting range of $T$ for the deconfinement transition. At finite temperature, the heavy $W$-bosons, of mass $\sim {1\over L}$, resulting from the $SU(3) \rightarrow U(1)^2$ breaking, can be excited with probability given by the Boltzmann factor, $\sim e^{- {m_W \over T}} \sim e^{- {{\cal{O}}(1) \over LT}}$, and their effect on the dynamics cannot be neglected. Since in the range (\ref{temperaturerange}) $m_W \sim {1\over L} \gg T$, the $W$-bosons are essentially static and form a non-relativistic 2d gas of electrically charged particles. Thus, one arrives \cite{Anber:2011gn} at the representation of the partition function of the thermal gauge theory as a 2d grand canonical neutral Coulomb gas of magnetically and electrically charged particles---the magnetic bions and the $W$-bosons. The ``particles"\footnote{To explain the use quotation marks: the electric objects, the $W$-bosons, are genuine particles, while the magnetic bions are instantons. Both appear as particles in the finite-$T$ dimensionally reduced description.} in each---magnetic or electric---component of the gas have their respective mutual (dual) Coulomb interactions, while the ``particles" belonging to different components  have  Aharonov-Bohm phase interactions with each other. This two-component Coulomb gas of electric and magnetic charges is equivalent to a generalization of the sine-Gordon model (see Section \ref{conclusions}).

A qualitative picture of the phase transition is as follows. At low-$T$, very few $W$-bosons are present in the magnetic bion plasma, so the confinement of electric charges in the magnetic-bion plasma persists, albeit with some amount of screening of the string tension. In this, $T<T_c$, phase the few $W$-bosons present are confined in neutral dipole pairs. As $T$ increases,  more dipole pairs appear, causing further screening of the confining string tension and thus an increase of the size of the dipoles, until, at $T_c$, a transition to a deconfined phase occurs---one where $W$-bosons form an electric plasma and the magnetic bions are confined in magnetically neutral dipole pairs, whose density and size, in turn, decrease  further as $T$ increases. We note that this picture is confirmed by our simulations---see the plot of the vortex density as a function of temperature on Fig.~\ref{fig:Vortex} (vortices in the spin model are identified with $W$-bosons).

Many---but not all---aspects of this picture can be studied analytically in the case of $SU(2)$ QCD(adj) via the renormalization group equations for the corresponding electric-magnetic Coulomb gas. In particular, various critical exponents can be exactly calculated for QCD(adj) with $SU(2)$ gauge group, due to the special properties of the appropriate electric-magnetic Coulomb gas  \cite{Anber:2011gn}. The gas turns out  to be dual to  a XY-spin model with a $\Z_4$-preserving perturbation and the phase transition in the spin model is in the same universality class as the deconfinement transition in $SU(2)$ QCD(adj). 

In the case of $SU(3)$ QCD(adj), a Coulomb gas picture of the theory and a spin-model dual to the Coulomb gas also exist.  However, the nature of the small-$L$ deconfinement transition for  $SU(3)$ gauge group is not well understood, as opposed to the $SU(2)$ case, as we now describe.
The results from  \cite{Anber:2011gn}    provide some guidance. An  important property of the $SU(3)$ QCD(adj) theory\footnote{Electric-magnetic duality in the Coulomb gases relevant for QCD(adj) on $\R^{2} \times \S^1_L \times \S^1_\beta$   does not hold for $SU(N_c\ge 4)$ due to the composite nature of the magnetic bions \cite{Anber:2011gn}. We will not study $N_c>3$ here.} on $\R^2 \times \S^1_\beta \times \S^1_L$   is that it exhibits electric-magnetic duality (valid in the temperature range (\ref{temperaturerange})). The duality interchanges electric and magnetic fugacities as well as the electric and magnetic coupling ($\kappa$ or $\kappa^{-1}$) and takes $\kappa \rightarrow {12\over \kappa}$ \cite{Anber:2011gn}. By eqn.~(\ref{kappasu3}) of Section \ref{natural}, this is also a high-$T$/low-$T$ duality,  suggesting that if there is a single transition,  the self-dual point is the critical point.
At present, we do not know if at the self-dual point (\ref{z3sqrdmodel}) is described by a conformal field theory   (this would be the case if the transition was continuous and is exactly what happens in the calculable case of  $SU(2)$). 
In \cite{Anber:2011gn}, it was shown that the leading order (in fugacities) renormalization group equation for $\kappa$ do, indeed, exhibit a fixed point at the self-dual point of the electric-magnetic duality.
 However, both $W$-boson and bion fugacities are relevant at the self-dual point---indicating that a strong-coupling analysis is necessary,  even for small fugacities  at the UV-cutoff.  Thus, determining the order of the transition calls for strong coupling methods---in particular, the present work.

 In Sections \ref{natural} and \ref{simulation}, we  describe the duality map of $SU(3)/\Z_3$ QCD(adj) on $\R^2 \times \S_L^1 \times S_\beta^1$ 
to an ``affine" XY-spin model. The discussion is meant to only   introduce the reader to  the duality and we only make what we consider the most relevant points. For more details, we refer to \cite{Anber:2011gn}.

\subsubsection{Spin model and symmetries in ``natural" (root-lattice) coordinates}
\label{natural}

The spin model dual to $SU(3)/\Z_3$ QCD(adj) is the theory of two  coupled XY-spins. The most natural description of the model is in terms of    
 two compact variables whose periodicity is determined by the $SU(3)$ root lattice:
 \begin{equation}
 \vec\theta_x =  (\theta^1_x, \theta^2_x) \equiv \vec\theta_x +  2 \pi  \vec\alpha_1 \equiv  \vec\theta_x +  2 \pi  \vec\alpha_2 ~.
\label{thetadef}
 \end{equation}
 Here $ \vec\alpha_{1,2}$ are the simple  roots\footnote{\label{weights}For future use, we give the weights of the defining representation:
$  \vec\nu_1= (\half, \frac{1}{2 \sqrt 3}), \; \vec\nu_2= (-\half, \frac{1}{2 \sqrt 3}),   \; 
\vec\nu_3= (0,  -\frac{1}{ \sqrt 3})$. The roots are differences of weights and are given by  
 $\vec\alpha_1 =  \vec\nu_1 -  \vec\nu_2=(1,0)$,   $\vec\alpha_2=  \vec\nu_2 -  \vec\nu_3=(-\half, \frac{\sqrt 3}{2})$, and affine root  $\vec\alpha_3=  \vec\nu_3 -  \vec\nu_1= (-\half, -\frac{\sqrt 3}{2})$. The weights obey $\vec \nu_i\cdot \vec\nu_j=\frac{\delta_{ij}}{2}-\frac{1}{6}$ for $i,j=1,2,3$.}  and $\vec\alpha_3$, which appears in (\ref{z3sqrdmodel}) below, is the affine root of $SU(3)$. In this Section, we 
 use root-lattice coordinates, as they provide  a natural description of the allowed electric and magnetic charges, the discrete symmetries, and the operators representing various nonperturbative objects in the gauge theory. 
 The affine XY-model is defined by a lattice partition function with:
\begin{equation}
\label{z3sqrdmodel}
- \beta H =   \sum_{x; \hat\mu = 1,2}  \sum_{i=1}^{N_c=3} {\kappa \over 4 \pi} \cos  2  \vec\nu_i \cdot (  \vec\theta_{x  + \hat\mu} - \vec\theta_{x}) + \sum_{x} \sum_{i=1}^{N_c=3}  \tilde{y}   \cos 2 ( \vec\alpha_i - \vec\alpha_{i-1})\cdot \vec\theta_{x} .
\end{equation}
where $\vec\nu_i$ $(i = 1,2,3)$ are the weights of the defining representation,  explicitly defined in Footnote \ref{weights}, $x$ denotes the sites on a 2d square lattice, and $\hat\mu$---the two unit lattice vectors.

The first term in (\ref{z3sqrdmodel}) is essentially  the model used by D.~Nelson  \cite{Nelson} to describe the melting of a two-dimensional crystal with a triangular lattice. There, the  two-vector $\vec{\theta}_x$ parametrizes fluctuations of the positions of atoms in a 2d crystal around equilibrium (the ``distortion field") and $\kappa$ is proportional to the Lam\' e coefficients. A vortex of the compact vector field $\vec{\theta}_x$ describes a dislocation in the 2d crystal. The Burger's vector of the dislocation is the winding number of the vortex, now also a two-component vector. For a description of the  order parameters and phase diagram, see \cite{Nelson}. We only mention that the melting of the 2d crystal occurs due to the proliferation of dislocations (vortices) at high temperature, which destroys the algebraic long-range translational order.

Now, we turn to the $SU(3)/\Z_3$ QCD(adj) meaning of the spin model.

The ``spin waves" in (\ref{z3sqrdmodel})---the fluctuations of $\vec\theta_x$---are the duals of the two massless photons. The dual photons are sourced by magnetically charged excitations---the magnetic bions, whose image in the spin model is the potential term in (\ref{z3sqrdmodel}), discussed below.

The  vortices in the spin model (\ref{z3sqrdmodel})  are dual to magnetic charges and thus
 describe the electric excitations in the gauge theory---the $W$-bosons of mass $\sim{1\over L}$, which can be excited at $T>0$, as described in the previous Section. There are three kinds of vortices of equal lowest action that contribute to the dynamics. These are described in \cite{Anber:2011gn}---and, much earlier, in \cite{Nelson}---and correspond to the three lightest degenerate $W$-bosons in $SU(3)$ QCD(adj) at small-$L$.  
The spin-spin (as well as vortex-vortex) coupling $\kappa$, which determines the kinetic term of the dual photons, is expressed via the four-dimensional gauge coupling $g_4(L)$, the size of the spatial circle $L$, and the temperature $T$:
\begin{equation}
\label{kappasu3}
\kappa = {1\over T} {g_4^2(L)\over 2 \pi L }. 
\end{equation}

 In the melting applications of the ``affine" XY-model, there is an exact 
$U(1)\times U(1)$ global symmetry $\vec{\theta}_x$$\rightarrow$$\vec{\theta}_x + \vec{c}$. The $U(1)^2$ symmetry forbids terms in the action which are not periodic functions of the difference operator. In the gauge theory, this is  the topological $U(1)^2$ symmetry of the two free dual photons. This symmetry is explicitly broken by the fact that the gauge theory possesses magnetic monopole-instanton solutions (essentially  't Hooft-Polyakov monopoles or ``twists" \cite{Lee:1997vp}
 thereof).

In the spin model dual to QCD(adj), the $U(1)^2$ symmetry is explicitly broken,  by the potential ``external field" term in (\ref{z3sqrdmodel}), to  a $\Z_3 \times \Z_3$ symmetry, to be defined below.  
This breaking of $U(1)^2$ is crucial (and is the main difference between (\ref{z3sqrdmodel}) and the models describing melting) as it  captures the effect of the magnetically charged excitations in the thermal gauge theory, the magnetic bions responsible for confinement. 
The strength of the  ``external field" in (\ref{z3sqrdmodel}), $\tilde{y}$,  is  proportional to the bion fugacity in the gauge theory.
The bion fugacity, see \cite{Anber:2011gn,Anber:2011de, Argyres:2012ka}, is 
 small at small $L$, and is given by (up to a numerical constant): 
\begin{equation}
\label{bionfugacity}
\tilde{y}  \sim {1\over T} {1 \over L^3 g^{14- 8 n_f}} \;e^{- S_0}~,
\end{equation}
 where $S_0 \simeq {8 \pi^2 \over g_4^2(L)}$   is the bion action.
In $SU(3)$, there are three kinds of magnetic  bions of different magnetic charges ($\sim$$\vec\alpha_i - \vec\alpha_j$, $i \ne j$) but  equal fugacities. The Coulomb gas of these three kinds of bions is represented in the spin model by the three equal-strength ``external field" terms in (\ref{z3sqrdmodel}). Thus, the model has an additional  symmetry, permuting the arguments of the three cosines, simultaneously in the kinetic and potential terms. This symmetry is inherited from the unbroken Weyl group  in the center-symmetric vacuum responsible for the  $SU(3) \rightarrow U(1)^2$ breaking. This symmetry remains unbroken at zero temperature and  we do not expect it to break at $T>0$ (indeed, the results of our simulations confirm this expectation, as the two magnetizations $m^{1}$ and $m^2$ (\ref{orderparameter}) and their susceptibilities  behave identically, but would not be expected to if the exchange symmetry between different bions and $W$ bosons was broken).  

As described in Section \ref{review},    in the temperature range (\ref{temperaturerange}) the gauge theory  has  an electric-magnetic Coulomb gas description as a gas of magnetic bions and $W$-bosons---both come in three varieties---interacting via electric and magnetic Coulomb potentials and Aharonov-Bohm interactions.
A Coulomb gas description also holds  for the spin model   and is the 
 reason behind the duality; see \cite{Jose:1977gm} for a lattice derivation and \cite{Anber:2011gn} for a continuum description.

The two $\Z_3$ symmetries\footnote{The reader not familiar with the Lie-algebraic constructions used in this Section is advised to proceed to Section \ref{simulation}, where a description in terms of coordinates rectifying the root lattice is given. While the physical interpretation of the two $\Z_3$ symmetries is not obvious in the  ``simulation" coordinates of Section \ref{simulation}, the presence of two $\Z_3$ symmetries is evident. } of the spin model (\ref{z3sqrdmodel}) are mapped to the topological $\Z_3^t$ of $SU(3)/\Z_3$ QCD(adj), associated with the nontrivial $\pi_1(SU(3)/\Z_3)$, and the $\Z_3^{d\chi}$ discrete subgroup of the anomaly-free global chiral symmetry. Note that while the $n_f$-adjoint flavor theory has a larger chiral symmetry group,  only $\Z_3^{d \chi}$ is relevant for the deconfinement transition on $\R^{1,2}\times \S_L^1$; the continuous $SU(n_f)$ chiral symmetry remains unbroken at small $L$. 
The topological\footnote{In \cite{Argyres:2012k a} this symmetry is called ``dual center symmetry", while earlier both ``topological global $\Z_N$ symmetry" \cite{'tHooft:1977hy} and ``magnetic $\Z_N$ symmetry" \cite{KorthalsAltes:2000gs} have been used.} $\Z_3^t$ acts on $\vec\theta$ by shifts on the weight lattice:\footnote{Despite appearances, there is only one $\Z_3$ symmetry in (\ref{z3t})---the difference between transformations with different $i$  is a shift of $\vec\theta$ by $2 \pi k$ times a root vector, which is an identification, as per (\ref{thetadef}).}
\begin{equation}
\label{z3t}
\Z_3^t: ~~ \vec\theta \rightarrow \vec\theta - 2 \pi k  \vec\nu_i,~ k = 0,1,2; ~ {\rm for ~any} ~ i = 1,2,3~. 
\end{equation}
The action of the other symmetry, the discrete chiral $\Z_3^{d \chi}$, is:
\begin{equation}
\label{z3dchi}
\Z_3^{d\chi}: ~~ \vec\theta \rightarrow \vec\theta + {2 \pi q \over 3}\vec{\rho}~, ~ q = 0, 1, 2~,
\end{equation}
where $\vec\rho$ is the Weyl vector\footnote{\label{weyl}The Weyl vector is defined as $\vec\rho =\vec\mu_1 + \vec\mu_2$, where $\vec\mu_i$ are the fundamental weights implicitly defined via $\vec\alpha_i \cdot \vec\mu_j = {\delta_{ij} \over 2}$. In the basis of Footnote \ref{weights}, $\vec\rho=({1 \over 2},{\sqrt{3} \over 2})$.} of $SU(3)$.
It is useful to describe the action of the two symmetries on appropriate order parameters. For the $\Z_3^t$ symmetry (\ref{z3t}),  the exponentials  $e^{2 i \vec\theta \cdot \vec\nu_j}$   can be taken as order parameters:
\begin{equation}
\label{orderparameters1}
 \Z_3^t: ~~ e^{2 i \vec\theta \cdot \vec\nu_j} \rightarrow e^{i {2 \pi k \over 3}} e^{2 i \vec\theta \cdot\nu_j}~.
\end{equation}
To obtain the result on the r.h.s., one uses the definitions from Footnote \ref{weights}.
An insertion of $e^{2 i \vec\theta_x \cdot \vec\nu_j}$ in the path integral of the spin-model (\ref{z3sqrdmodel}) 
corresponds to the creation, at $x$, of a pointlike monopole of minimal charge   allowed by Dirac quantization in an $SU(3)/\Z_3$ gauge theory. Recall that in $SU(N_c)/\Z_{N_c}$ theories,  dynamical 't Hooft-Polyakov monopoles have magnetic charges taking values in the root lattice. Thus, their charges are larger  than the minimal ones allowed by Dirac quantization (the  minimal allowed magnetic charges take values in the finer weight lattice).\footnote{See Sect.~2.2 in \cite{Argyres:2012ka} for a rather formal (but valid for general simple Lie groups) discussion in the  $\R^{1,3} \times \S^1_L$ context. Alternatively,  Sect.~2.2.1 in \cite{Anber:2011gn} contains a less Lie-algebra loaded  explanation for $SO(3)$ and $SU(2)$.} As  explained in   \cite{Anber:2011gn,Argyres:2012ka}, the operator (\ref{orderparameters1}), when lifted  to the gauge theory,  creates a minimal charge 't Hooft loop at  $x \in \R^2$,  wrapping around $\S^1_L$. 

In the $SU(3)/\Z_3$ gauge theory, no electric probes in the  fundamental representation are allowed. Thus, there is no center symmetry and the symmetry\footnote{See \cite{Smilga:1996cm} for discussions of  ``center symmetry breaking at high-$T$", $\Z_N$ bubbles,  and  spatial 't Hooft loops.} associated with confinement is  the ``dual center" $\Z_3^t$ symmetry associated with $\pi_1(SU(3)/\Z_3)$. Confinement is thus probed by the 
behavior of correlators of minimal-charge monopoles.  At low temperatures, the  $\Z_3^t$ symmetry is broken and the correlator of two separated  minimal-charge monopoles (\ref{orderparameters1}) (or  't Hooft loops around $\S_L^1$)  should approach unity at large separation. Conversely, in the deconfined phase, the correlator has an exponential fall off and approaches zero at large distances, owing to the confinement of minimal charge monopoles at high-$T$.

Under the discrete chiral $\Z_3^{d \chi}$ symmetry (\ref{z3dchi}),  the operators creating dynamical magnetic monopole-instantons---'t Hooft-Polyakov monopoles with charges  in the   root lattice---transform as:
\begin{equation}
\label{orderparameters2}
\Z_3^{d\chi}: ~~ e^{2 i \vec\theta \cdot \vec\alpha_j} \rightarrow e^{i {2 \pi q \over 3}} e^{2 i \vec\theta \cdot\alpha_j}~,
\end{equation} 
where  (\ref{z3dchi}) and Footnotes \ref{weights} and \ref{weyl} were used to obtain the r.h.s.~(it is easily seen that the operators (\ref{orderparameters2}) are invariant under $\Z_3^{t}$).
Monopole-instantons are  charged under the discrete chiral symmetry due to the intertwining of the topological shift symmetries and discrete chiral symmetries. This intertwining occurs due to the presence of  fermion zero-modes in a monopole-instanton background, required by the Nye-Singer index theorem \cite{Nye:2000eg}. Note that magnetic bions have no fermion zero modes and are chiral-symmetry neutral, hence the external field term in the spin model (\ref{z3sqrdmodel}) representing bions is neutral under (\ref{z3dchi}). 
At zero temperature, the $\Z_3^{d \chi}$ symmetry is broken, hence, the correlator of two monopole operators (\ref{orderparameters2})  approaches unity at large separation. We expect  this symmetry to be restored at high $T$. In fact, we will see that---as was also found \cite{Anber:2011gn} in the $SU(2)$ theory---the restoration of the topological and discrete chiral symmetries occurs at the same temperature, as we see only a single transition. 

Finally, we note that operators representing electrically charged objects---the $W$-bosons---can also be defined in the spin model as ``vortex creation operators" \cite{Anber:2011gn}. We have not simulated correlators of such operators---both because they are not local in the  $\vec\theta_x$ basis (and thus require significant more computer resources to study) and because  they are not charged under any global symmetry of the $SU(3)/\Z_3$ theory---and will not discuss them here.

\subsubsection{Spin model and symmetries in ``simulation" coordinates}
\label{simulation}
%%%%%%%%%%%%%%%%%%%%%%%%%%%%%%%%%%%
In the simulation, we find it more convenient to use coordinates that rectify the root lattice.   
An equivalent description of (\ref{z3sqrdmodel}) is given  in terms of two compact scalar fields on a 2d square lattice, $\phi^i_x \equiv \phi^i_x + 2\pi$ $(i = 1,2)$, which are both periodic by $2\pi$. These are related to $\vec\theta_x$ as follows:
\begin{equation}
\label{rectification}
\phi^1_x = - \theta^1_x +   {1 \over \sqrt{3}} \; \theta^2_x~,~~   
\phi^2_x = - \theta^1_x -  {1 \over \sqrt{3}}\; \theta^2_x~.
\end{equation}
 The action of the lattice model (\ref{z3sqrdmodel}) becomes, using $\nabla_{\mu}\phi_{ {x}} = \phi_{ {x}+\hat{\mu}}-\phi_{{x}}$:
\begin{equation}\label{action}
\begin{split}
	-\beta\mathcal{H} &= \frac{\kappa}{4\pi}\sum_{{x};\hat{\mu}=\hat{1},\hat{2}} \left(\cos(\nabla_{\mu}\phi_{{x}}^1) + \cos(\nabla_{\mu}\phi_{{x}}^2) + \cos(\nabla_{\mu}\phi_{{x}}^1 - \nabla_{\mu}\phi_{{x}}^2)\right)  \\
& \quad + \tilde{y}\sum_{{x}}\left( \cos(3\phi_{{x}}^1)+\cos(3\phi_{{x}}^2) + \cos(3(\phi_{{x}}^1-\phi_{{x}}^2)) \right)~.
\end{split}
\end{equation}
 We now  define a  ``temperature" $T^\prime$, by ${1 \over T^\prime}\equiv {\kappa \over 2 \pi}$, proportional to the gauge theory temperature, see (\ref{kappasu3}), as well an ``external field strength" $h = {2 \pi \tilde{y} \over \kappa}$, see (\ref{bionfugacity}). The lattice spacing of the spin model is taken to be equal to unity (and, physically, is of order $L$, the inverse UV cutoff scale of the 2d Coulomb gas). 
Thus, we arrive at the Boltzmann factor which is actually evaluated at every iteration of the Metropolis algorithm employed in our simulations:
\begin{eqnarray}
\label{boltzmann}
e^{-\beta\mathcal{H}} &\equiv & e^{-{H\over T'}} \nonumber \\
H &\equiv &
   -\; \frac{1}{2}\sum\limits_{{x};\hat{\mu}=\hat{1},\hat{2}} 
   \left(\cos(\nabla_{\mu}\phi_{ {x}}^1) + \cos(\nabla_{\mu}\phi_{ {x}}^2) + \cos(\nabla_{\mu}\phi_{ {x}}^1 - \nabla_{\mu}\phi_{ {x}}^2)\right)  \\
 && -\; h\;\sum_{{x}}\left( \cos(3\phi_{ {x}}^1) +\cos(3\phi_{ {x}}^2)  + \cos 3(\phi_{ {x}}^1-\phi_{{x}}^2)  \right) ~.\nonumber
\end{eqnarray} 
The action (\ref{boltzmann}) has a $\Z_3^1\times \Z_3^2$ symmetry, $\Z_3^1$ acting solely on $\phi^1$ by a shift of $2 \pi \over 3$ times integer, and $\Z_3^2$, acting only on $\phi^2$ by a ${2 \pi  \over 3}$ shift. Using the map (\ref{rectification}), it is easy to see that the topological $\Z_3^t$ maps to a linear combination of $\Z_3^{1}$ and $\Z_3^2$:
\begin{eqnarray}
\label{z3tsimulation}
\Z_3^t: \phi^1 \rightarrow \phi^1 + {2 \pi k \over 3}, ~~ \phi^2 \rightarrow \phi^2 - {2 \pi k \over 3}~, ~k = 0,1,2, 
\end{eqnarray}
while the chiral $\Z_3^{d \chi}$ is identical to $\Z_3^2$:
\begin{eqnarray}
\label{z3chiralsimulation}
\Z_3^{d\chi}: \phi^1 \rightarrow \phi^1, ~~ \phi^2 \rightarrow \phi^2 + {2 \pi q \over 3}~,~ q=0,1,2.
\end{eqnarray}

The main goal in this paper is to study the symmetry realization as the temperature $T^\prime$ and coupling $h$ in (\ref{boltzmann}) vary. Below, we define the quantities measured in our simulations. 
The ``magnetization",  $M^j$, for each field component ($j=1,2$) is the average $j$-th ``spin", a planar unit vector represented as $e^{i \phi_x^j}$, over the entire lattice: 
\begin{equation}\label{magnetization}
	 M ^j = \sum_{ {x}} e^{i \phi^j_{ {x}}}~,~ ~ j=1,2.
\end{equation}
The order parameter  for the $\Z_3^j$ symmetry is the mean magnetization $m^j$ per spin:
\begin{equation}\label{orderparameter}
	m^j = \frac{1}{N^2} \langle  \big\vert \sum_x e^{i \phi_{ {x}}^j} \big\vert \rangle = \frac{\langle |M^j|\rangle}{N^2}~,~ ~ j=1,2.
	\end{equation}
	Here and in Section \ref{mcresults} we use $N$ to denote the lattice width (the number of lattice points is $N^2$).
We also define the magnetic susceptibility $\chi^j$ of each component:
\begin{equation}\label{chi}
	\chi^j = \frac{\langle | {M}^j|^2 \rangle - \langle| {M}^j|\rangle^2}{N^2T'} ~,~ ~ j=1,2, 
\end{equation}
as well as the specific heat:
\begin{equation}\label{specificheat}
	C = \frac{\langle H^2\rangle - \langle H \rangle^2}{N^2T'^2}.
\end{equation}
We also measure the average energy per site (or average energy per spin, remembering that the spin is a two-component vector), defined as:
\begin{equation}\label{energyperspin}
E = {\langle H \rangle \over N^2}~.
\end{equation}

%%%%%%%%%%%%%%%%%%%%%%%%%%%%%%%%%%%%%%%%%%%%%%%%%%%
\section{Monte Carlo simulations}
%%%%%%%%%%%%%%%%%%%%%%%%%%%%%%%%%%%%%%%%%%%%%%%%%%%
\label{mcresults}

We have simulated  (\ref{boltzmann}) on a square lattice of widths $N =8, 16$, and $32$, and with perturbations  of strength $h=0.1$ and $1.0$ using the Metropolis algorithm. These values for $N$ and $h$ were used for most of the direct measurements. In our  study of the finite-size scaling of the susceptibility and specific heat we also used $h=0.5$ and added $N=24$ for all values of $h$.
 The data are obtained after $20000$ Monte Carlo sweeps (a sweep consists of $N^2$ Metropolis iterations), taking data every $10$ sweeps and disregarding the first $2000$ sweeps for equilibration. Data acquisition is performed either by starting from a random configuration at high temperature and then decreasing the temperature, or by starting from an ordered configuration at low temperature and then increasing  the temperature. We take the temperature difference $\Delta T'$ to be as small as $0.01$ in the critical region, and  $\Delta T' = 0.001$ in the case of finite-size scaling. Data points and error bars were generated from raw data using the bootstrap method\footnote{For succinct explanation of the bootstrap method, see \cite{Newman-Barkema}.} using 1000 resamples. 

\subsection{Direct measurements: magnetization, energy, and vortex density}\label{DirectMeasurements}

\begin{figure}[ht]
    {
    \parbox[c]{\textwidth}
        {
	\centering
%	\begin{subfigure}{0.55\textwidth}
  %      	\includegraphics[angle=0, width=\textwidth]{h030000M1.pdf}
%		\caption{$h=0$}
%	\end{subfigure}
	\begin{subfigure}{0.55\textwidth}
	\centering
        	\includegraphics[angle=0, width=\textwidth]{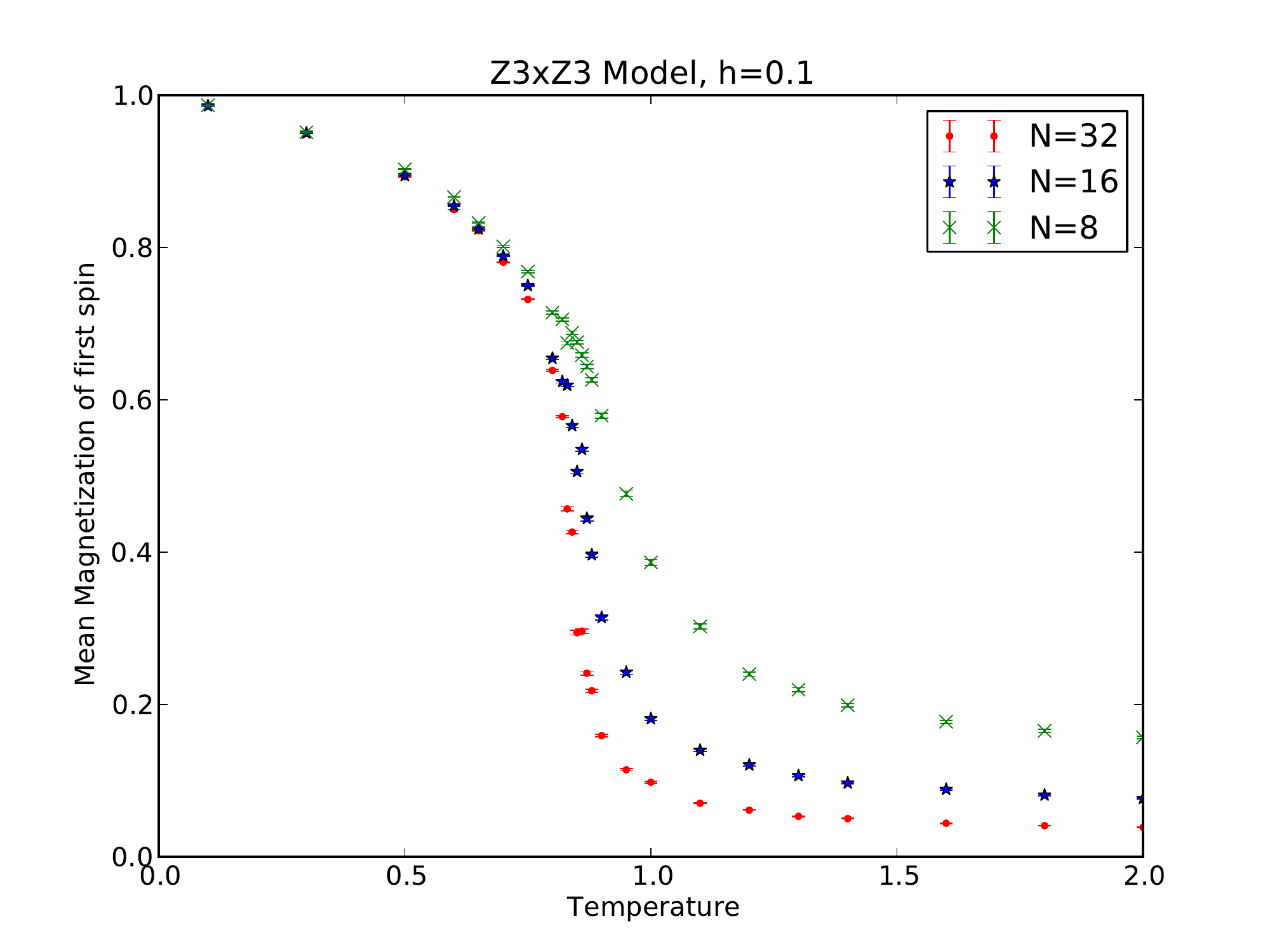}
		\caption{$h=0.1$}
	\end{subfigure}
 \centering
	\begin{subfigure}{0.551\textwidth}
	\centering
        	\includegraphics[angle=0, width=\textwidth]{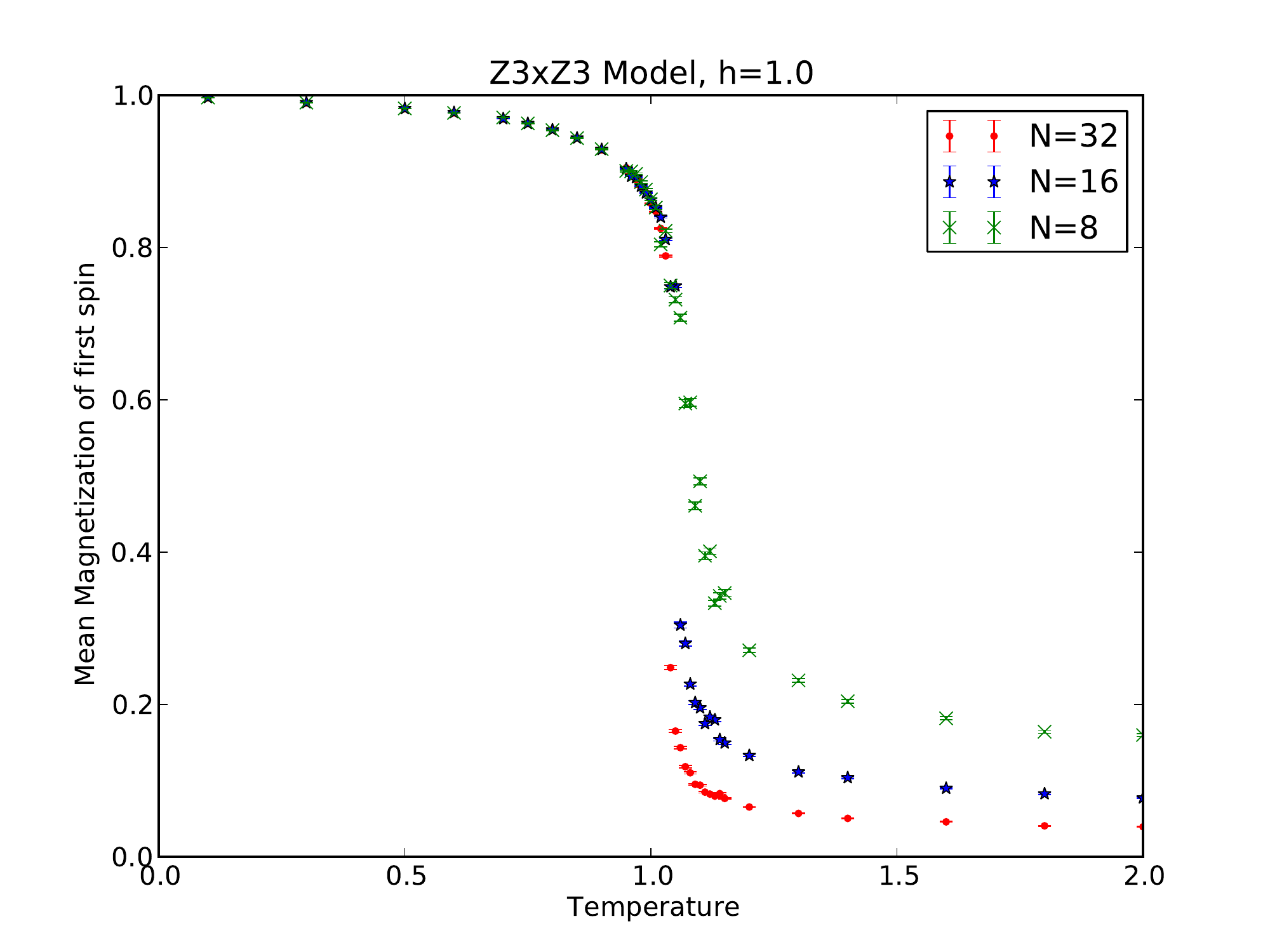}
		\caption{$h=1.0$}
	\end{subfigure}
	\hfil
	      \caption 
     	{ 
 	Order parameter vs. temperature $T'$ of the $\Z_3 \times  \Z_3$ model with symmetry-breaking fields of strength 
% {\rm (a)}: $h=0$, 
 {\rm (a)}: $h=0.1$, and {\rm (b)}:
 $h=1.0$ in lattices of width $N=8$ (green crosses), $N=16$ (blue stars), and $N=32$ (red circles).}
        \label{fig:Mag}
       % \end{center}
        }
      }
\end{figure}

We begin  by a discussion of the measurements of magnetizations, energy, and vortex density. 
The order parameter of the system (magnetization per spin, eq.~(\ref{orderparameter}))  is shown on 
Figure \ref{fig:Mag} as a function of the temperature  $T'$ for fugacities $h=0.1$ \& $1.0$, for lattices of width $N=8,16$ \& $32$. We found that the magnetizations   $m^1$ and $m^2$ behave identically and only show results for $m^1$.
 In all cases, the magnetization at high temperatures (beyond the observed transition) depends on the width of the lattice, with larger lattices more closely approaching the continuum limit of $m^i\rightarrow 0$ as $T'\rightarrow \infty$. The transition also appears to occur more sharply as $N$ increases (for a given $h$), and as the strength of the perturbation $h$ increases. 
 
 Upon running simulations with increasing and decreasing temperature, we have found no evidence of hysteresis. While this observation alone would favor a continuous transition, such a hypothesis is not consistent with our other findings discussed below (perhaps, the failure to detect hysteresis is due to a combination of: using rather small volumes, a large tunneling rate, and a sufficiently long simulation time).

\begin{figure}[ht]
    {
    \parbox[c]{\textwidth}
        {
%	\begin{subfigure}{0.55\textwidth}
  %      \centering
    %    \includegraphics[angle=0, width=\textwidth]{h030000U.pdf}
%	\caption{$h=0$}
%	\end{subfigure}
	\begin{subfigure}{0.55\textwidth}
	 \centering
        \includegraphics[angle=0, width=\textwidth]{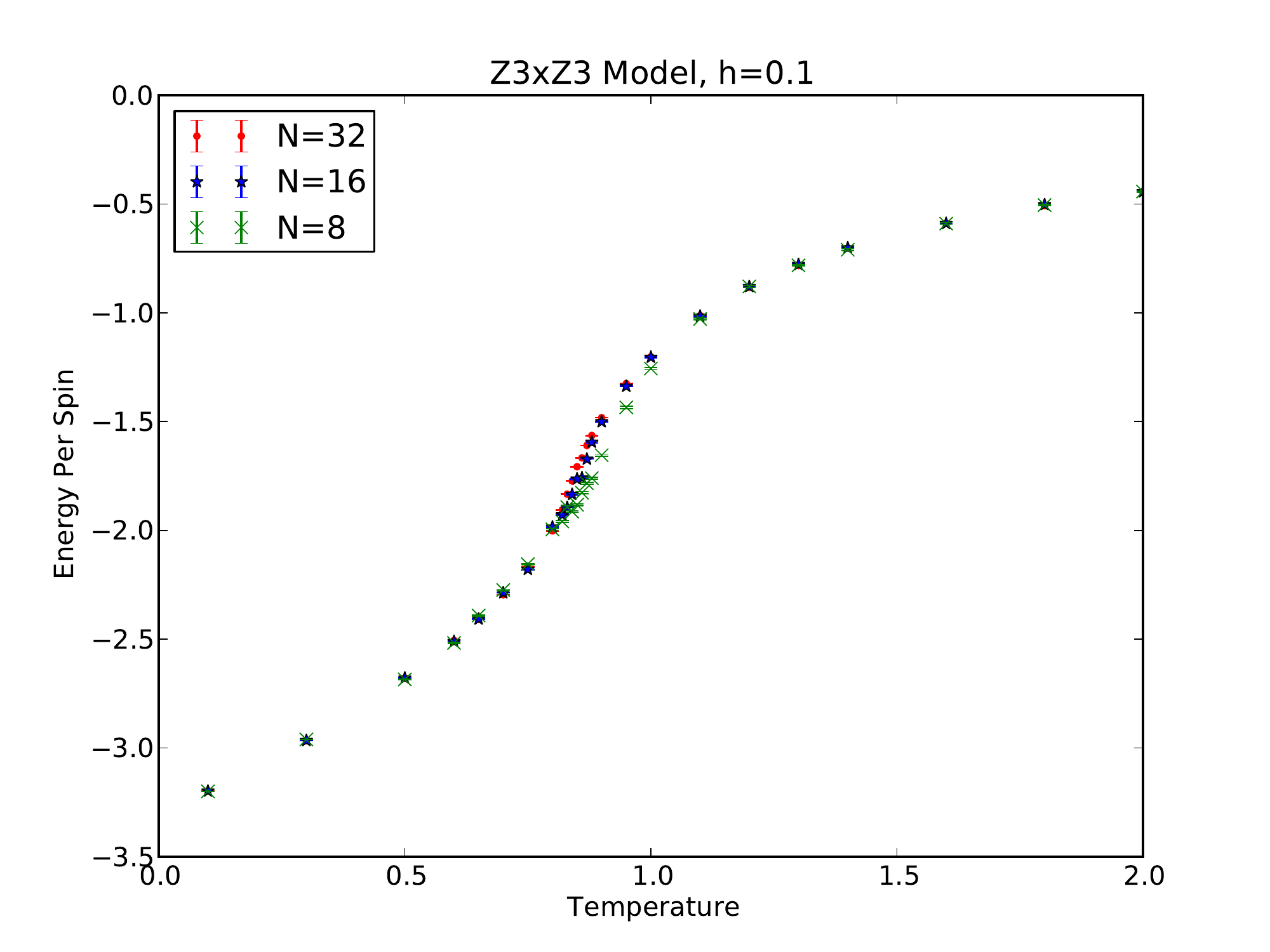}
	\caption{$h=0.1$}
	\end{subfigure}
	\centering
	\begin{subfigure}{0.55\textwidth}
        \centering
        \includegraphics[angle=0, width=\textwidth]{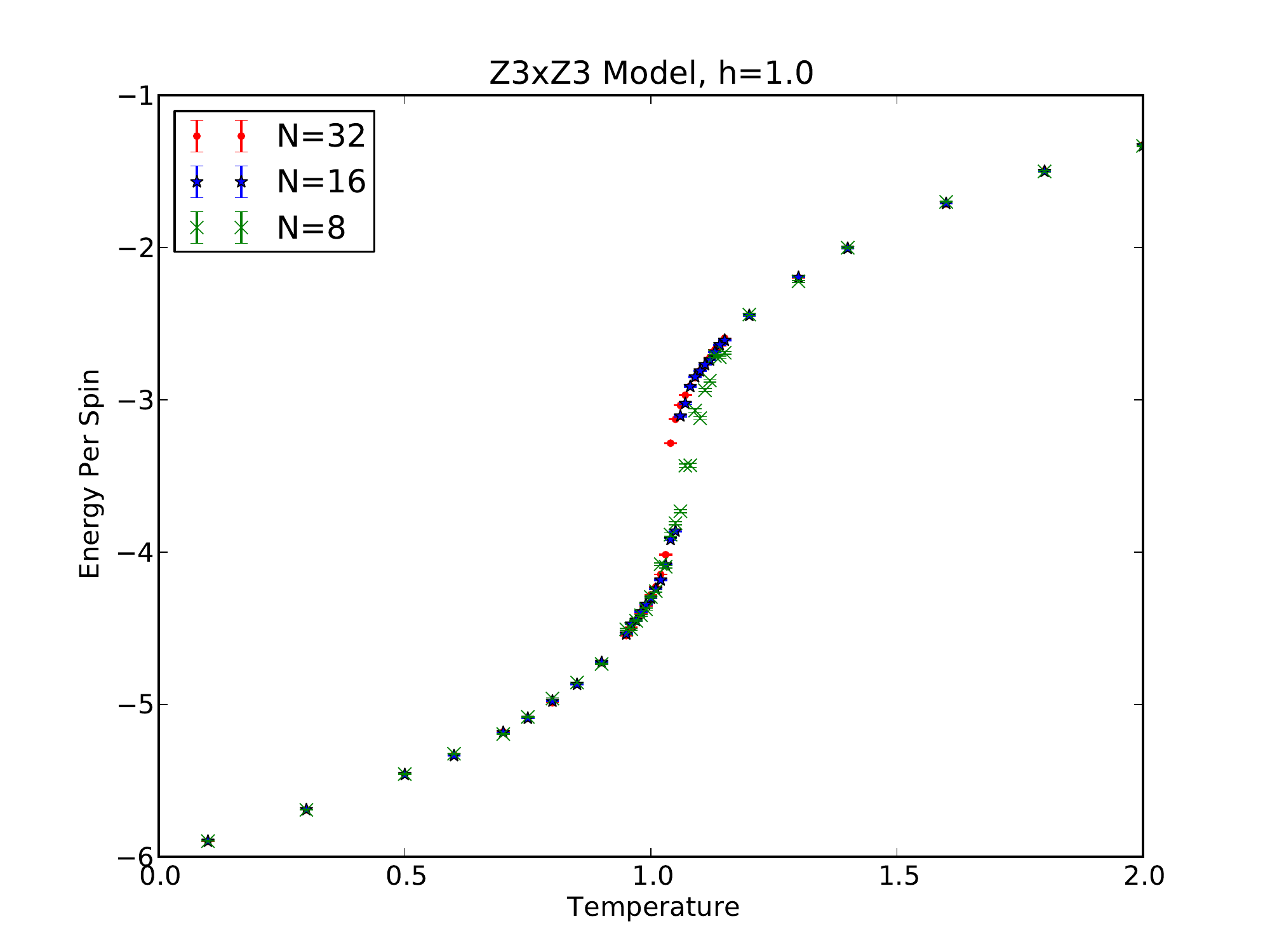}
	\caption{$h=1.0$}
	\end{subfigure}
	\hfil

        \caption 
      {  Energy per spin  vs. temperature $T'$ of the $\Z_3 \times  \Z_3$ model with symmetry-breaking fields of strength
      % {\rm (a)}: $h=0$,
       {\rm (a)}: $h=0.1$, and {\rm (b)}: $h=1.0$ in lattices of width $N=8$ (green crosses), $N=16$ (blue stars), and $N=32$ (red circles).     
			}
        \label{fig:Energy}
        %\end{center}
        }
    }
\end{figure}

The energy per spin (\ref{energyperspin}) vs. temperature is shown on 
Figure \ref{fig:Energy}   for the various strengths of $h$ and lattice widths. Similarly, the transition becomes sharper as $h$ increases. The observed low-temperature limits of $E=\frac{\langle H\rangle }{N^2} = -3.3$ \& $-6$ for $h= 0.1$ \& $1.0$, respectively, correspond exactly to the state where all spins are aligned, as expected from eqn.~(\ref{boltzmann}). 
 
 The vortex number was determined by calculating the change of the angle over each plaquette, from one site to the next,  making sure that the angle difference between neighboring sites lies in the range from $-\pi$ to $\pi$. The vortex density of each field was  recorded after every sweep. Vortices were detected with a sensitivity of $10^{-5}$, meaning that at each plaquette, see \cite{Tobochnik:1979zz}, a vortex of type\footnote{The density of $(1,0)$ vortices is identical to that of $(0,1)$ vortices; see \cite{Anber:2011gn} for an review of the  vortices in the multi-spin XY-model. The third  vortex  of the same fugacity, the $(1,1)$ vortex is difficult to distinguish from overlapping $(1,0)$ and $(0,1)$ vortices in the simulation.} $(1,0)$---i.e., a vortex of $\phi^1$ but not $\phi^2$---was detected if the sum of the difference in the angles $\phi^1$ of the spins around the plaquette's border was within $10^{-5}$ of $2\pi$ (higher-order vortices are assumed, and are, in fact, suppressed). 

\begin{figure}[ht]
    {
    \parbox[c]{\textwidth}
        {
%	\begin{subfigure}{0.55\textwidth}
   %     \centering
   %     \includegraphics[angle=0, width=\textwidth]{h030000Nu.pdf}
	%\caption{$h=0$}
	%\end{subfigure}
	\begin{subfigure}{0.55\textwidth}
	 \centering
        \includegraphics[angle=0, width=\textwidth]{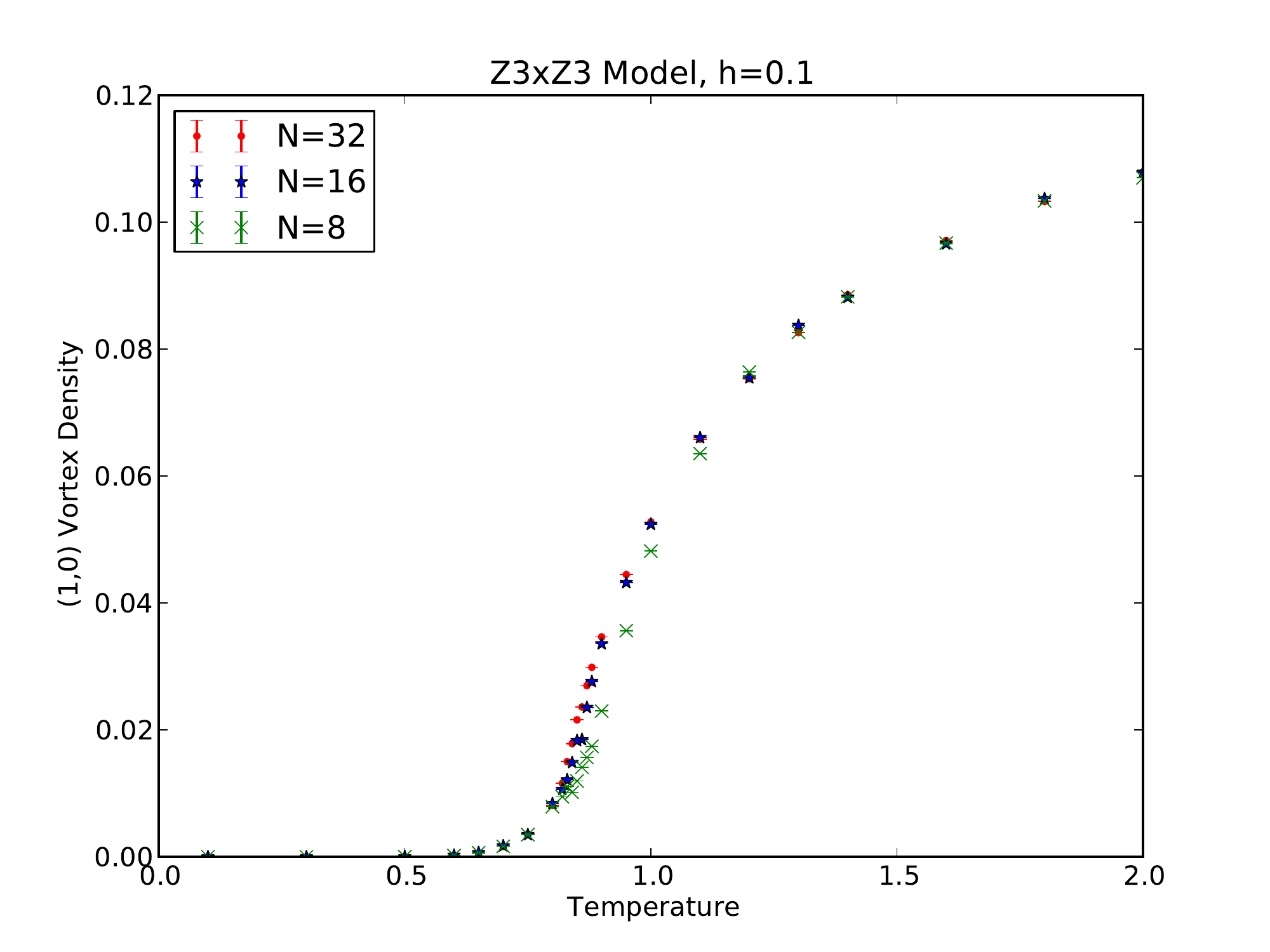}
	\caption{$h=0.1$}
	\end{subfigure}
	\begin{subfigure}{0.55\textwidth}
\centering
        \includegraphics[angle=0, width=\textwidth]{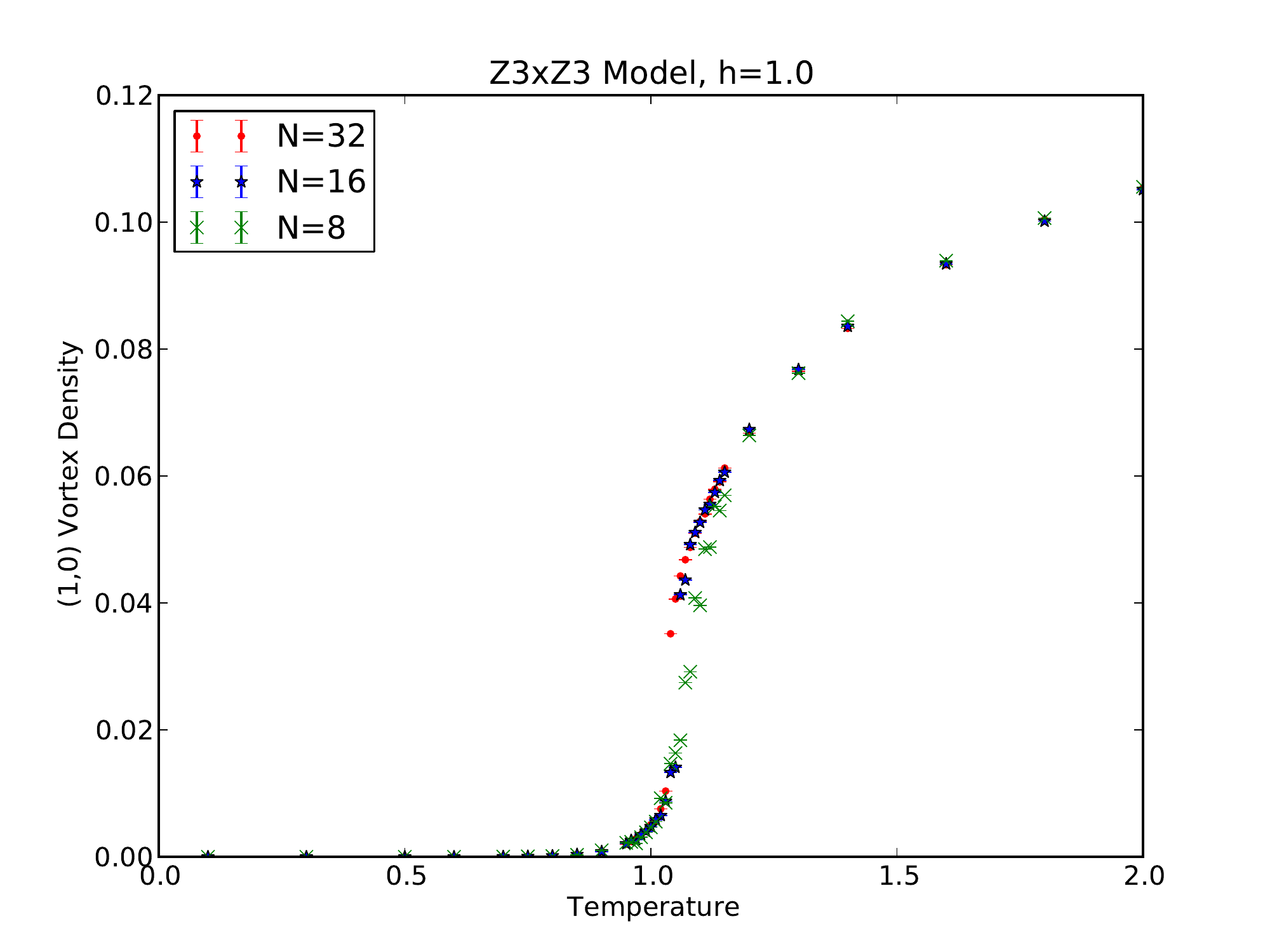}
	\caption{$h=1.0$}
	\end{subfigure}
	\hfil

        \caption 
      {   Vortex density vs. temperature $T'$ of the $\Z_3 \times  \Z_3$ model with symmetry-breaking fields of strength
      % {\rm (a)}: $h=0$,
        {\rm (a)}: $h=0.1$, and {\rm (b)}: $h=1.0$ in lattices of width $N=8$ (green crosses), $N=16$ (blue stars), and $N=32$ (red circles).  
			}
        \label{fig:Vortex}
        %\end{center}
        }
    }
\end{figure}

Figure \ref{fig:Vortex} shows the evolution of the $(1,0)$ vortex density as a function of temperature for the various strengths of perturbation and lattice widths. In all cases, the vortex density transitions from $0$ at low temperatures to about $0.11$ at temperatures above the critical temperature. Again, this transition is sharpest for $h=1.0$. The behavior of the vortex (i.e., $W^\pm$-boson) density is consistent with the interpretation of the deconfinement transition as occurring due to the ``liberation", upon increase of $T$, of the electric charges confined in small dipole pairs at low temperatures.

 \subsection{Magnetic susceptibility and specific heat}
\begin{figure}[ht]
    {
    \parbox[c]{\textwidth}
        {
	%\begin{subfigure}{0.55\textwidth}
        %\centering
        %\includegraphics[angle=0, width=\textwidth]{h030000Chi1.pdf}
%	\caption{$h=0$}
%	\end{subfigure}
	\begin{subfigure}{0.55\textwidth}
	 \centering
        \includegraphics[angle=0, width=\textwidth]{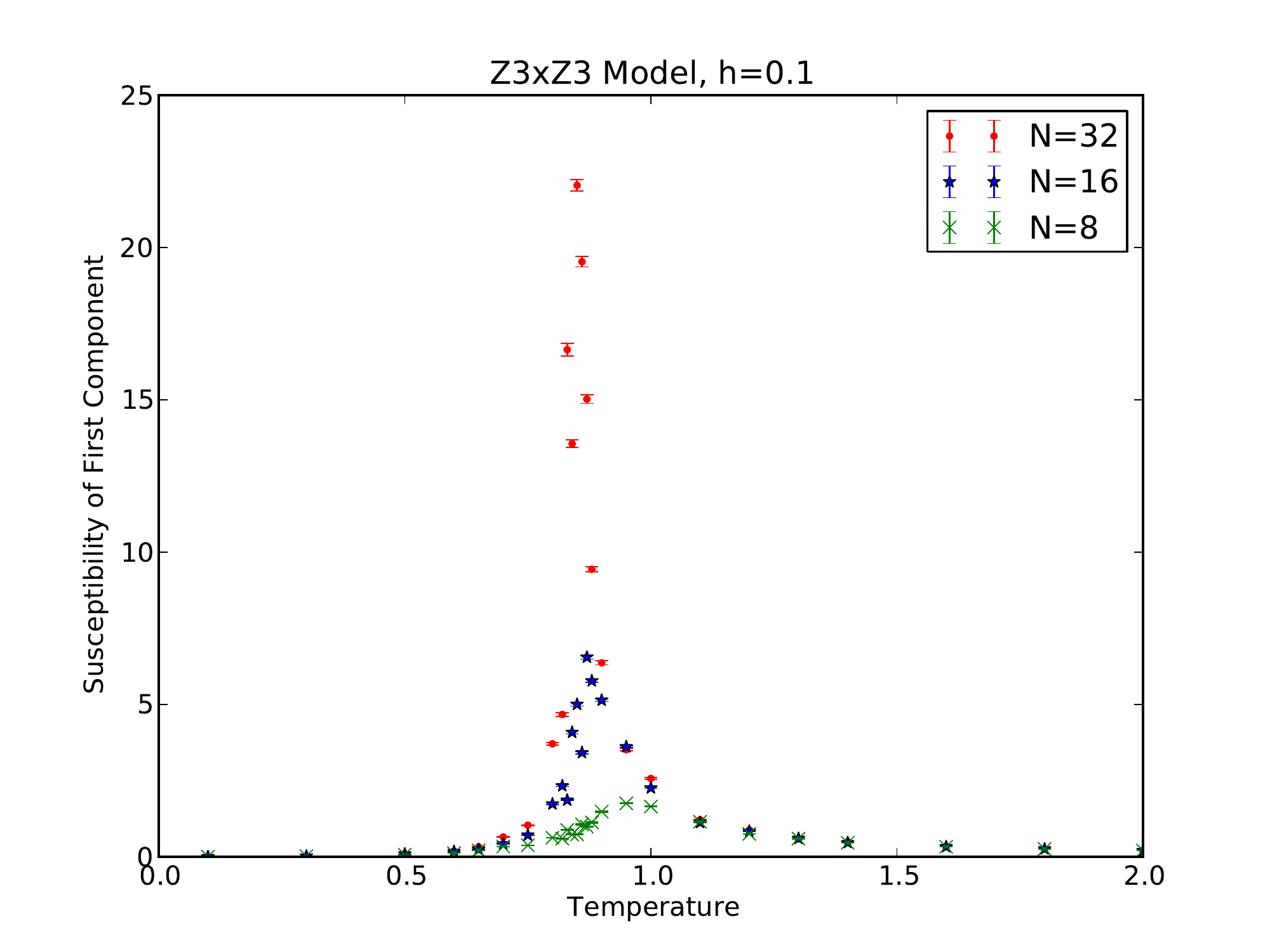}
	\caption{$h=0.1$}
	\end{subfigure}
	\begin{subfigure}{0.55\textwidth}
        \centering
        \includegraphics[angle=0, width=\textwidth]{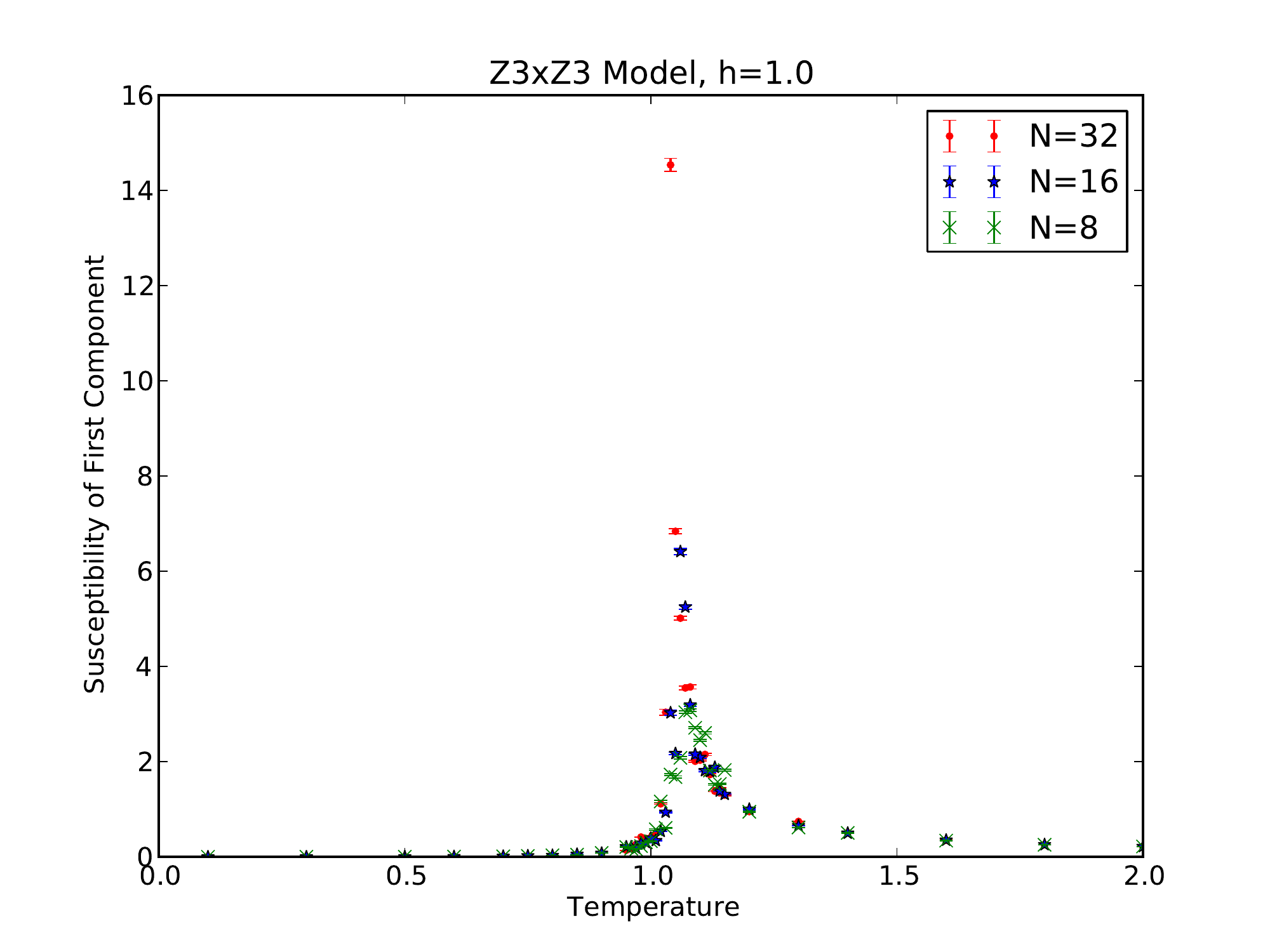}
	\caption{$h=1.0$}
	\end{subfigure}
	\hfil

        \caption 
      {  Magnetic susceptibility vs. temperature $T'$ of the $\Z_3 \times  \Z_3$ model with symmetry-breaking fields of strength 
      %{\rm (a)}: $h=0$, 
      {\rm (a)} $h=0.1$ and {\rm (b)} $h=1.0$ in lattices of width $N=8$ (green crosses), $N=16$ (blue stars), and $N=32$ (red circles).
			}
        \label{fig:Susceptibility}
        %\end{center}
        }
    }
\end{figure}

The magnetic susceptibility (\ref{chi}) as a function of temperature, for the same systems as discussed in Section~\ref{DirectMeasurements}, is shown on Figure \ref{fig:Susceptibility}. 
In all cases, the susceptibility is roughly $0$ at both high and low temperatures, and attains a peak around the critical region of temperature. The  height of the peak depends on the size of the lattice. In our simulations, the systems of lattice width $N=32$ reached the highest peaks in susceptibility. The location of these peaks provide an estimate of the   critical temperature. 

\begin{figure}[ht]
    {
    \parbox[c]{\textwidth}
        {
	%\begin{subfigure}{0.55\textwidth}
       % 	\centering
     %   	\includegraphics[angle=0, width=\textwidth]{h030000C.pdf}
%%	\end{subfigure}
	\begin{subfigure}{0.55\textwidth}
		 \centering
        	\includegraphics[angle=0, width=\textwidth]{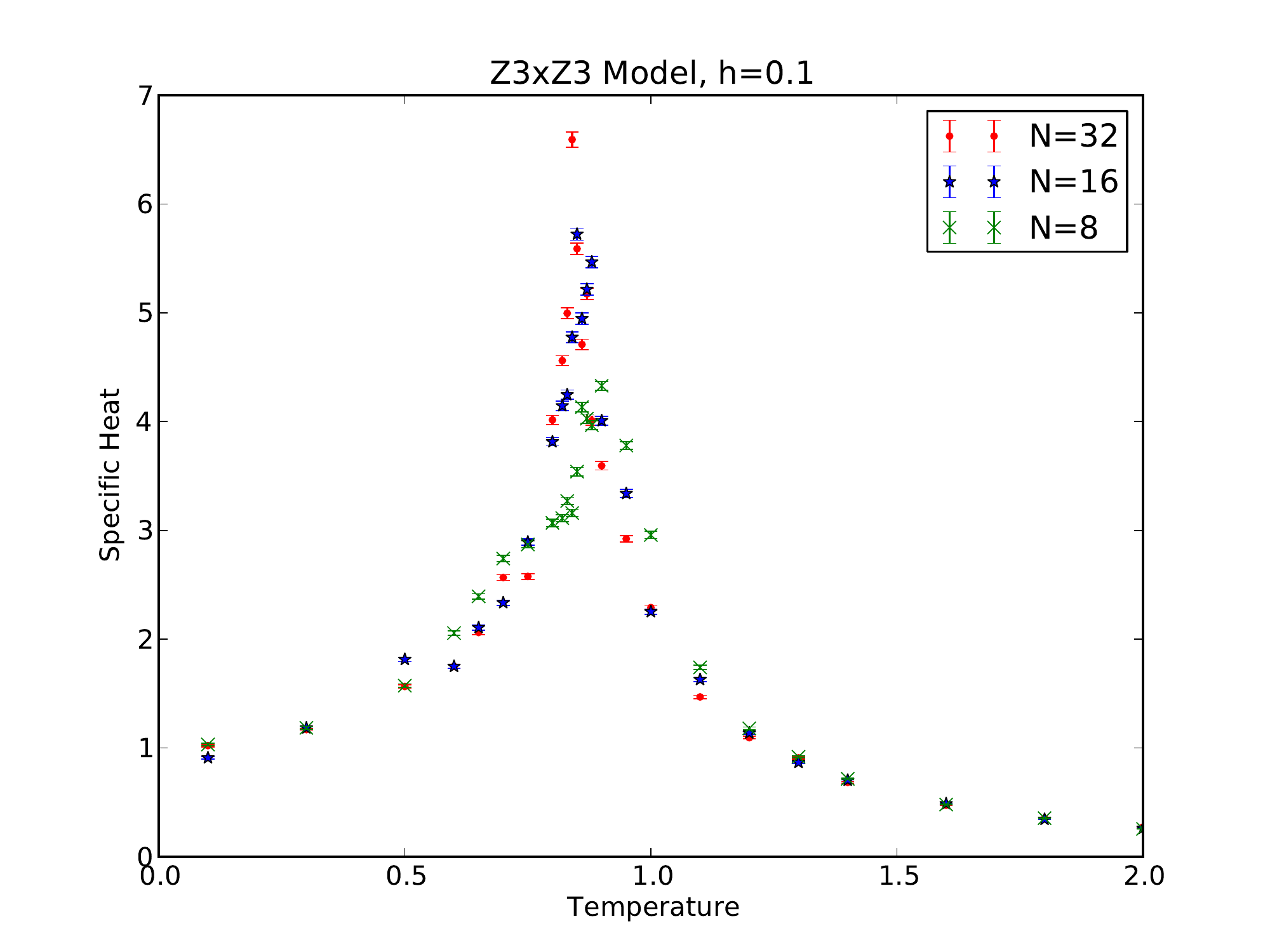}
		\caption{$h=0.1$}
	\end{subfigure}
	\begin{subfigure}{0.55\textwidth}
        	\centering
        	\includegraphics[angle=0, width=\textwidth]{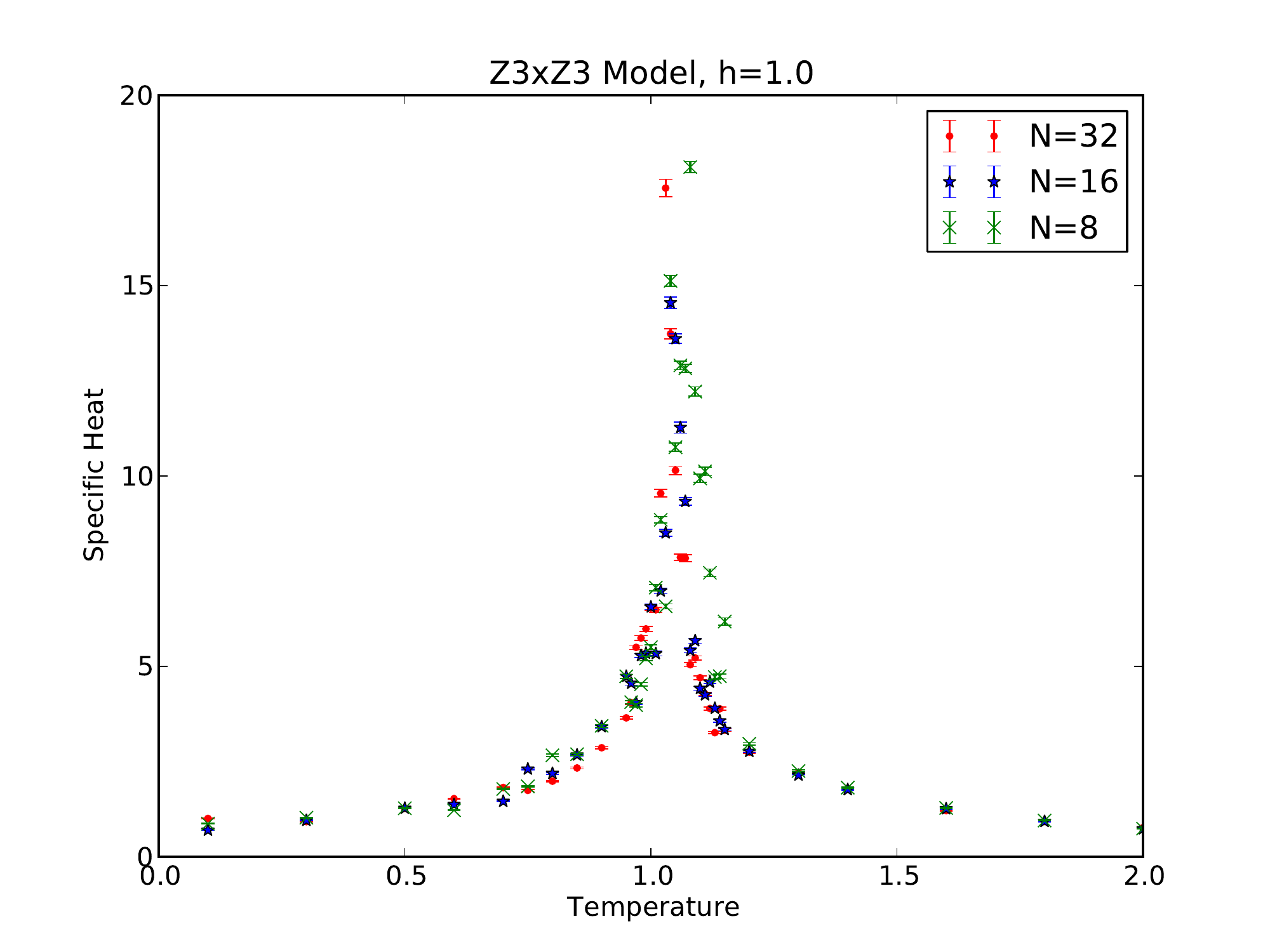}
		\caption{$h=1.0$}
	\end{subfigure}
	\hfil

        \caption 
      {   Specific heat vs. temperature $T'$ of the $\Z_3 \times  \Z_3$ model with symmetry-breaking fields of strength 
      %{\rm (a)}: $h=0$, 
      {\rm (a)} $h=0.1$ and {\rm (b)} $h=1.0$ in lattices of width $N=8$ (green crosses), $N=16$ (blue stars), and $N=32$ (red circles).
			}
        \label{fig:SpecificHeat}
        %\end{center}
        }
    }
\end{figure}

The specific heat (\ref{specificheat}) of the systems in question ($h=0.1,1$; $N=8,16,32$) is shown on 
 Figure \ref{fig:SpecificHeat}  as a function of temperature $T'$. The specific heat transitions from $C=1$ at low temperatures to a sharp peak in the critical temperature region and subsequently decreases at high temperatures. The height of the peak varies  with  lattice size and increases with the strength of the perturbation $h$. 
 We caution the reader against using the plots of this Section to infer details of the finite-size scaling---a more accurate study employing much finer $\Delta T'$ intervals in the critical region, as well as an additional lattice width and coupling, is presented in 
  the following Section.

\subsection{Finite-size scaling and energy probability distribution}
\label{finitesize}

To perform finite-size scaling, we studied the behavior of the magnetic susceptibility and the heat capacity in the critical region of each lattice width. We used  additional simulations with $\Delta T' = 0.001$ in the critical region  for each lattice size. In addition to the two cases $h=0.1$ and $h=1.0$, we also considered $h=0.5$. In summary, in our finite-size scaling study, four different lattice widths $N=8,16,24,32$ were used for all three values of $h=0.1,0.5,1$.

The best finite-size scaling fit we found is consistent with the hypothesis of a first-order deconfining phase transition. As explained in \cite{Challa:1986sk} and references therein, in a finite volume the delta-function like behavior of the susceptibility $\chi$ characteristic for  a first-order transition is smeared out to a peak scaling as $N^2$ (in $d=2)$ of width $\sim N^{-2}$. Similarly, the maximal value of the heat capacity $C$ increases like $N^2$. The maximal value of $\chi^i$ as a function of $T'$, $\chi_{max}(N)$ is expected to behave as:
\begin{equation}
\label{chimaxscale}
 \chi_{max}^i(N) = c_0 N^2 + \ldots~,
 \end{equation} 
 where dots denote terms subleading at large $N$.\footnote{If one uses the usual definitions of critical exponents in a first-order transition to characterize the finite-size scaling, the scaling in (\ref{chimaxscale}) would imply ${\gamma\over \nu} = d$.}
The maximum value of the specific heat as a function of $T'$ at each $N$, $C_{max}(N)$,  obeys:
  \begin{equation}
  \label{cmax}
  C_{max}(N) = {(E_+ - E_-)^2  \over 4 T_c'(\infty)^2} N^2 + \ldots~,
  \end{equation}
where $T'_c(\infty)$ is the infinite volume critical temperature   \cite{Challa:1986sk}. 
Here, $E_+$ and $E_-$ are the average energies per site of the symmetric ($E_+$) and broken ($E_-$) phase at $T_c'$. In particular, $E_+ - E_-$ is the latent heat (per site) of the transition. As the variation of $T'_c(N)$ with $N$ we find is slight (thermodynamic fluctuation theory predicts ${T'_c(N)-T'_c(\infty) \over T'_c(\infty)} \sim N^{-2}$),  we can   use the slope of the $C_{max}$ vs. $N^2$ line in (\ref{cmax}) (found in our relatively small volumes) to estimate the latent heat per spin.

\begin{figure}[ht]
    {
    \parbox[c]{\textwidth}
        {
 	\begin{subfigure}{0.46\textwidth}
         	\centering 
         	\includegraphics[angle=0, width=\textwidth]{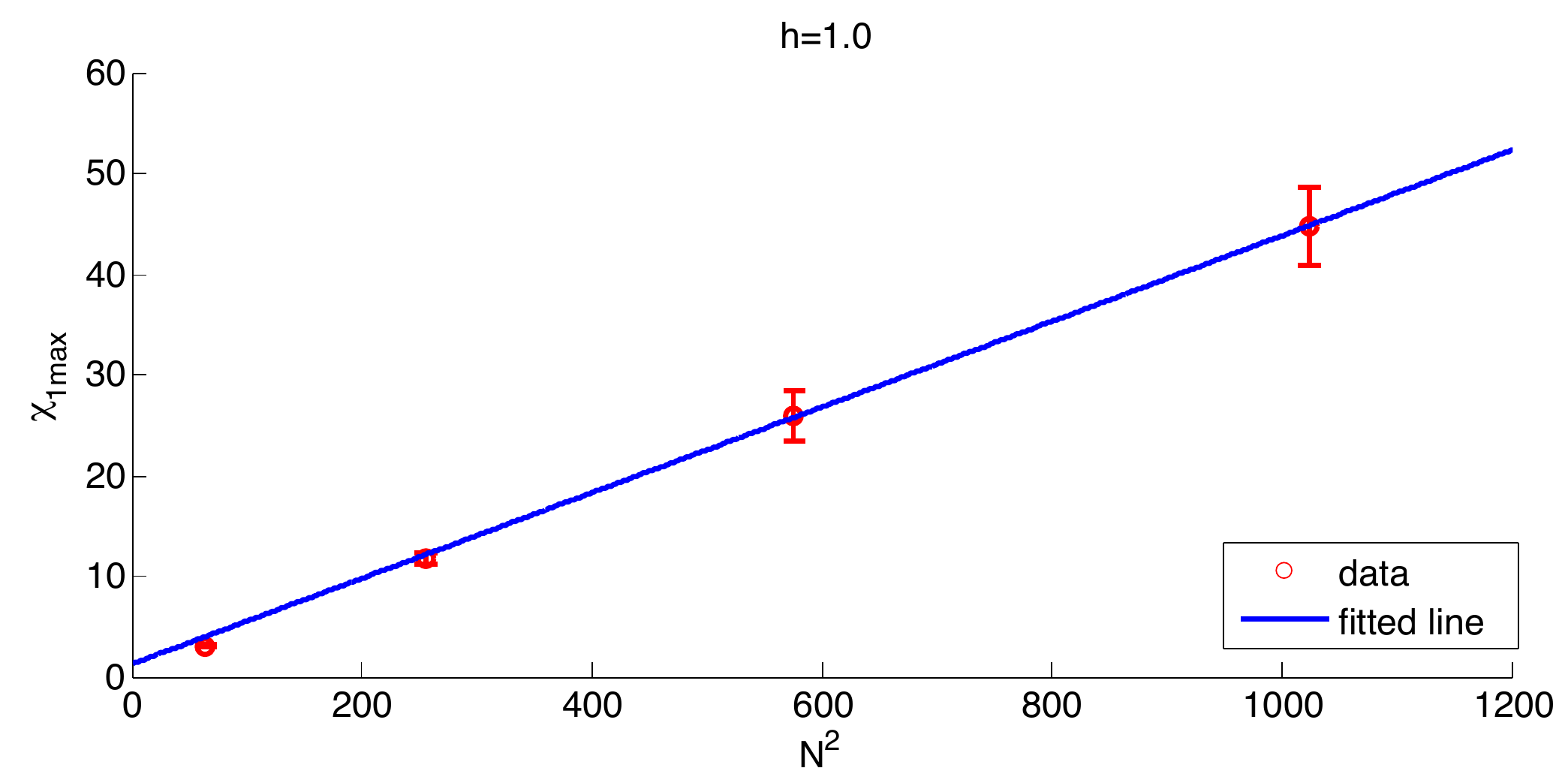}
 	\caption{   Maximum of the first component of susceptibility vs. $N^2$.}
 	\end{subfigure}
	\begin{subfigure}{0.46\textwidth}
		 \centering
        	\includegraphics[angle=0, width=\textwidth]{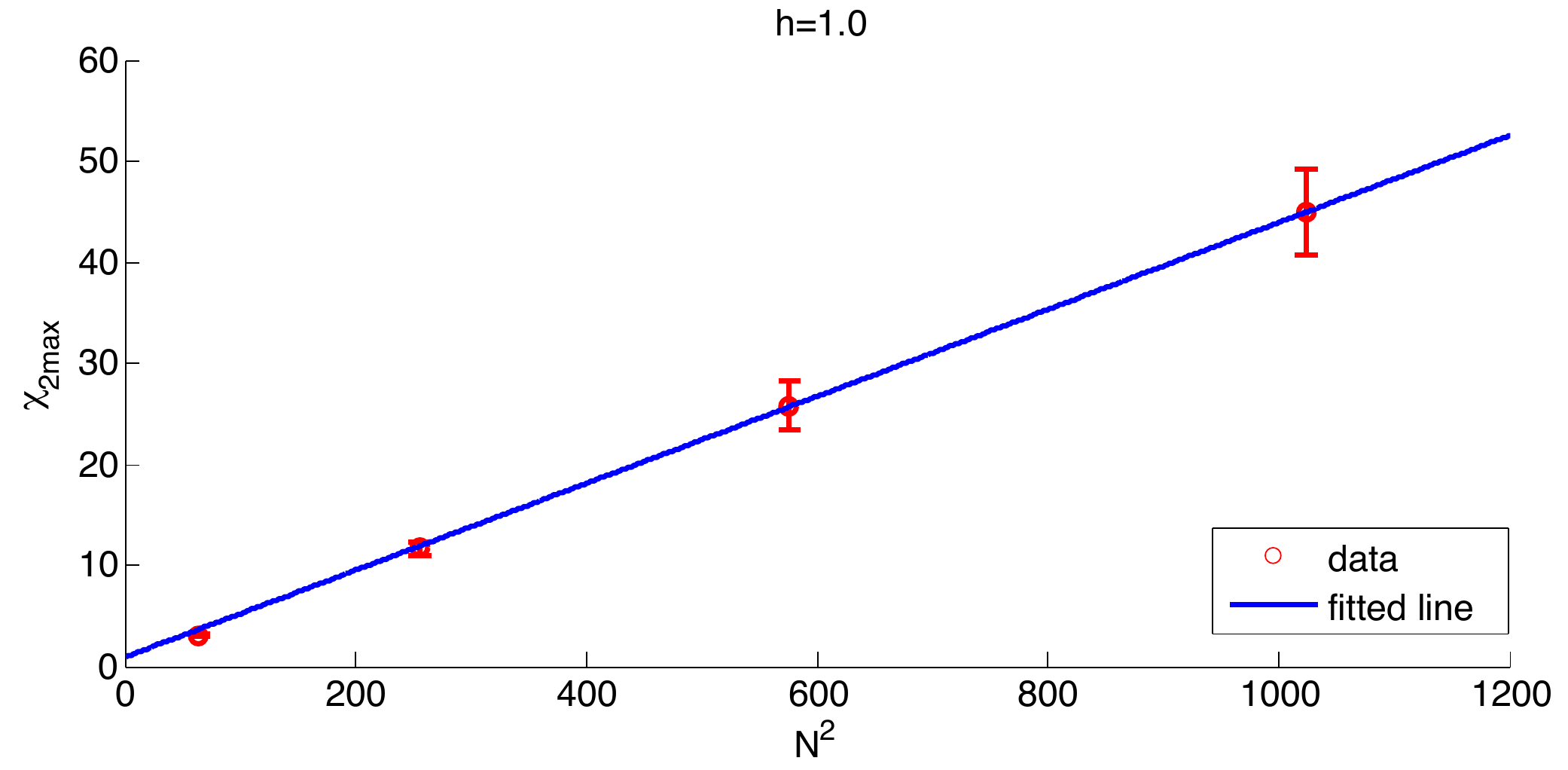}
		\caption{ Maximum of the second component of susceptibility vs. $N^2$.}
	\end{subfigure}
	\begin{subfigure}{0.46\textwidth}
        	\centering
        	\includegraphics[angle=0, width=\textwidth]{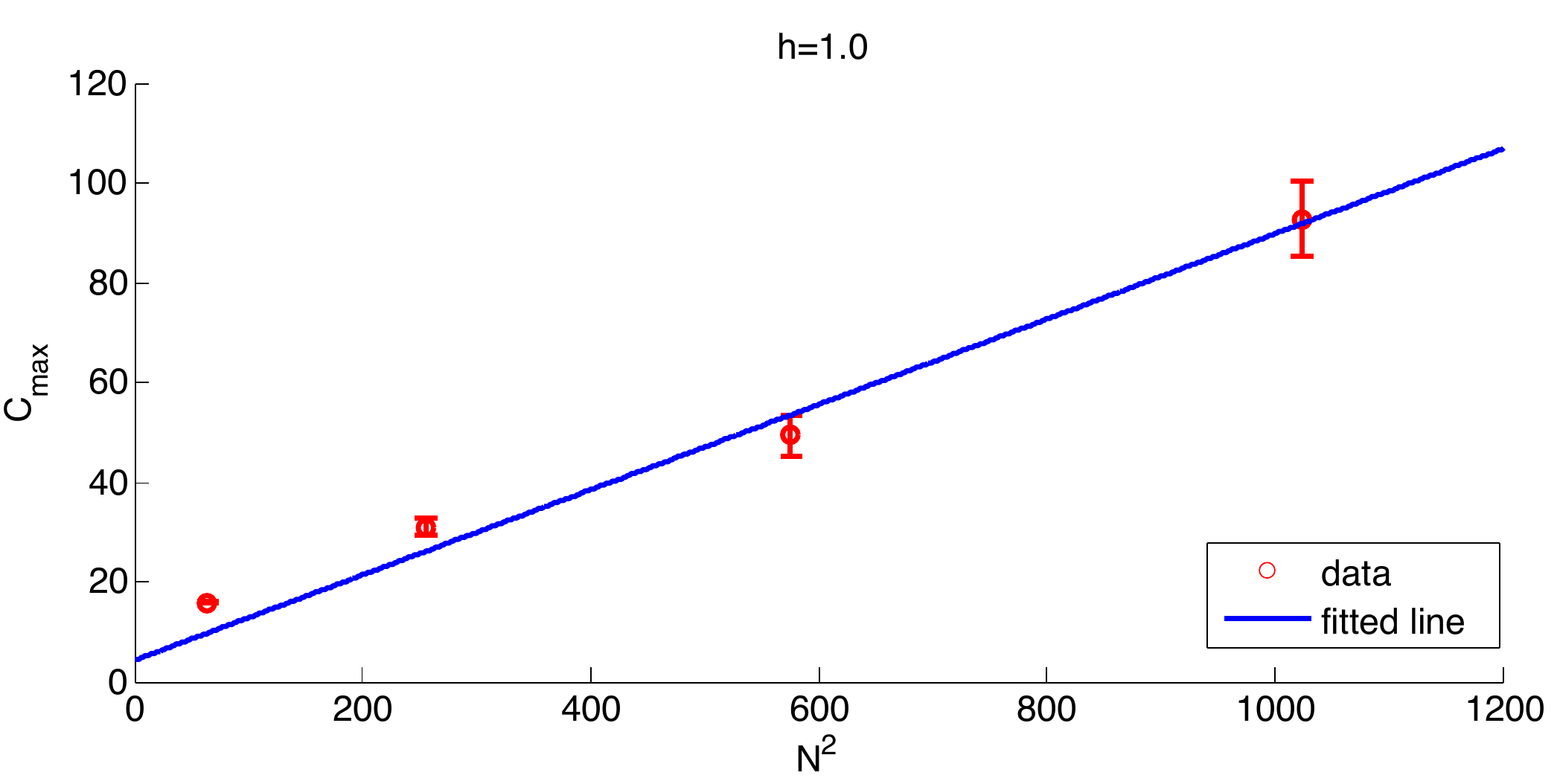}
		\caption{  Maximum of the specific heat per site vs. $N^2$.}
	\end{subfigure}
		\hfil
  \caption 
      {   Finite-size scaling of $\chi^i_{max}(N)$, $i=1,2$, and $C_{max}(N)$ for $h=1$.
			}
        \label{fig:h1FSS}
        %\end{center}
        }
    }
\end{figure}

\begin{figure}[ht]
    {
    \parbox[c]{\textwidth}
        {
 	\begin{subfigure}{0.46\textwidth}
         	\centering 
         	\includegraphics[angle=0, width=\textwidth]{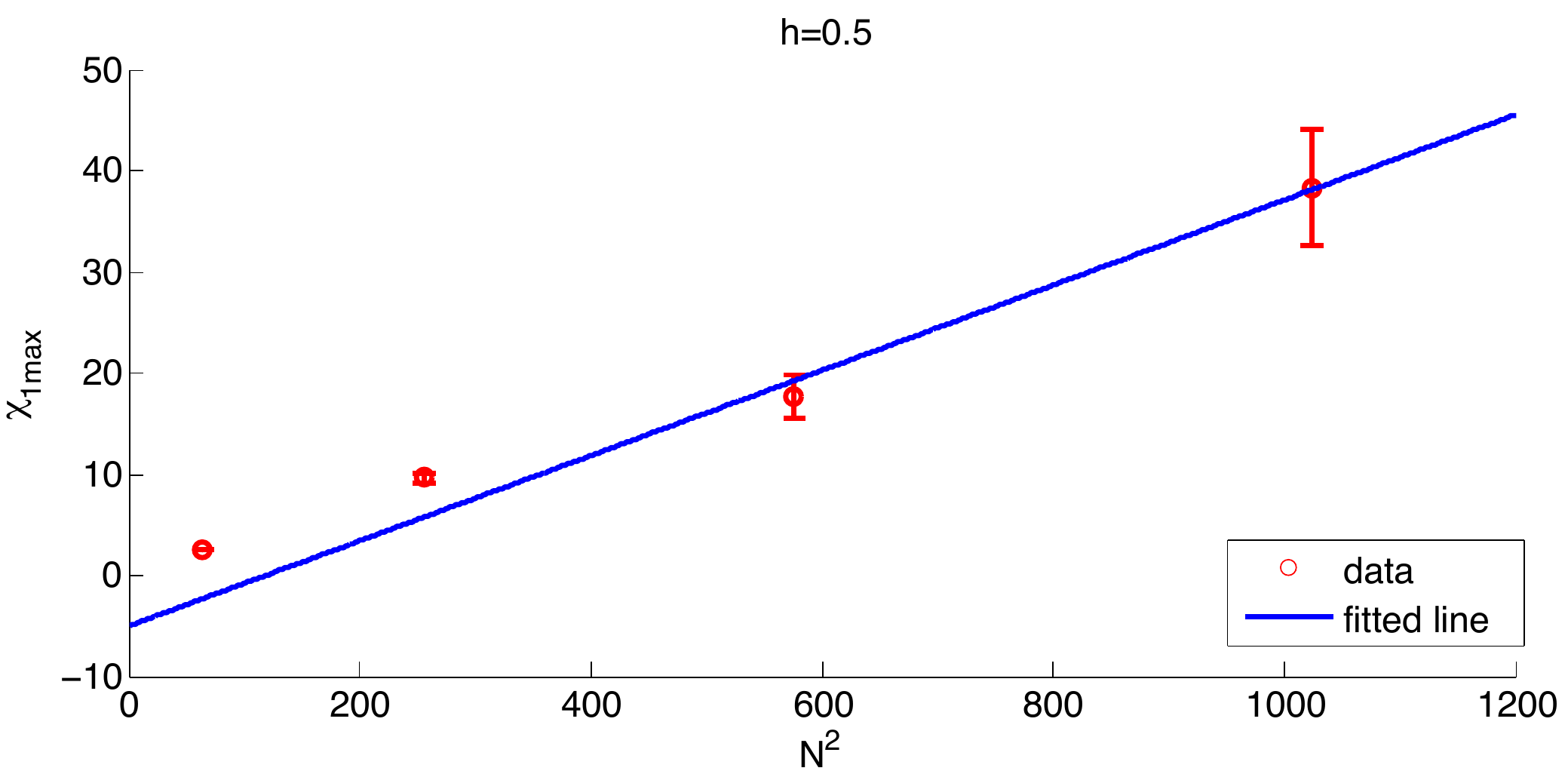}
 	\caption{ Maximum of the first component of susceptibility vs. $N^2$.}
 	\end{subfigure}
	\begin{subfigure}{0.46\textwidth}
		 \centering
        	\includegraphics[angle=0, width=\textwidth]{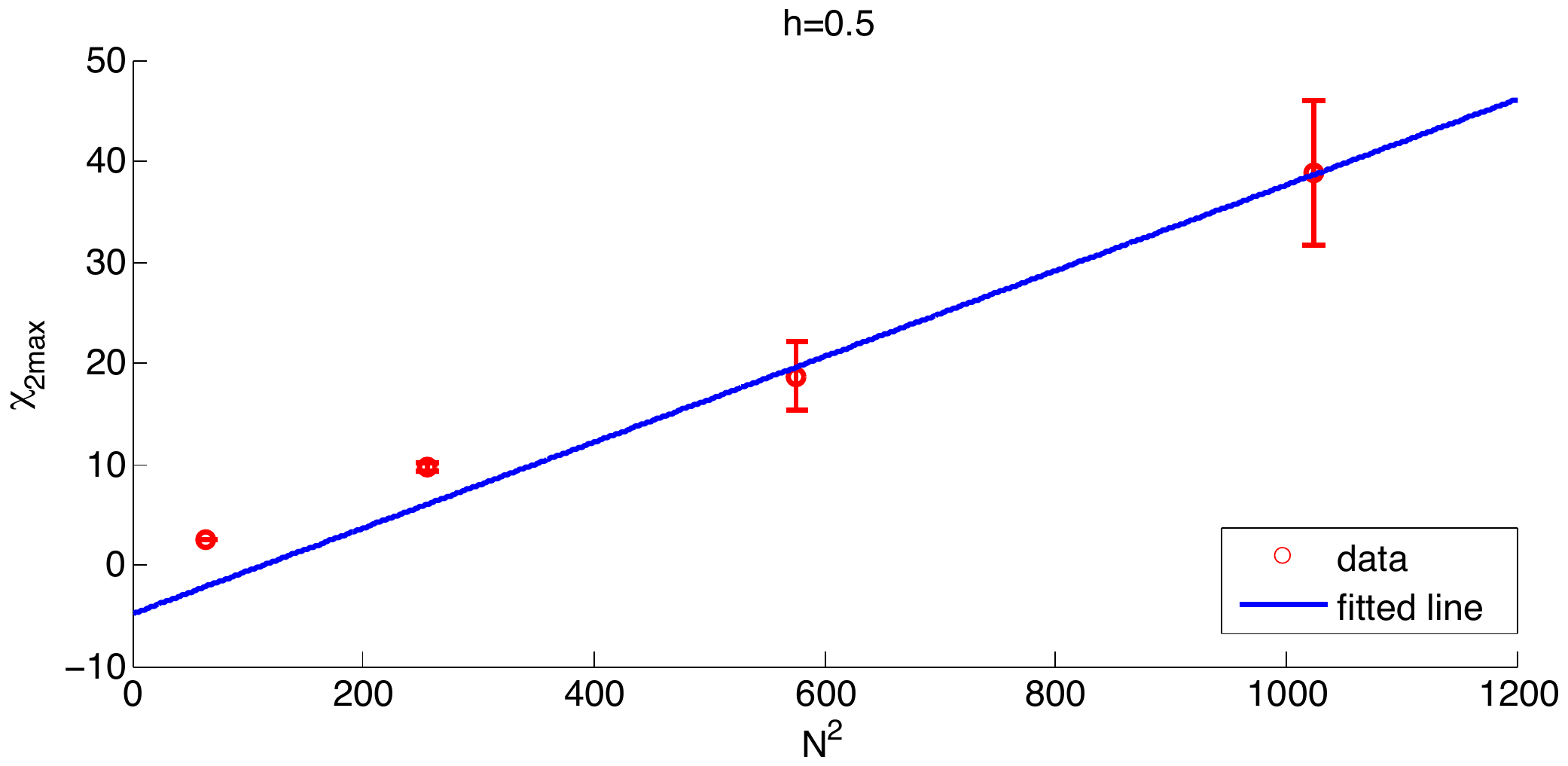}
		\caption{ Maximum of the second component of susceptibility vs. $N^2$.}
	\end{subfigure}
	\begin{subfigure}{0.46\textwidth}
        	\centering
        	\includegraphics[angle=0, width=\textwidth]{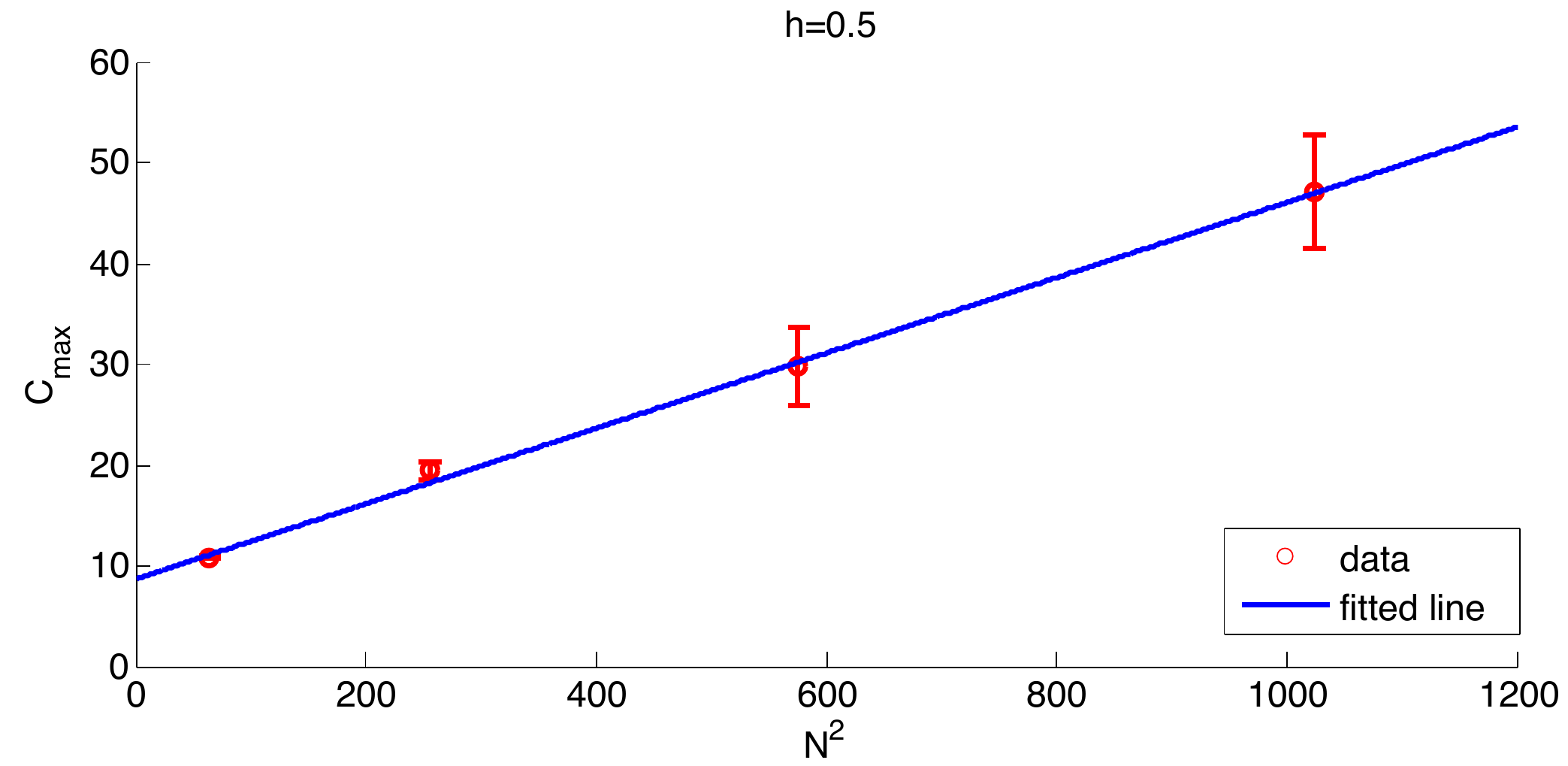}
		\caption{ Maximum of the specific heat per site vs. $N^2$.}
	\end{subfigure}
		\hfil
  \caption 
      {   Finite-size scaling of $\chi^i_{max}(N)$, $i=1,2$, and $C_{max}(N)$ for $h=0.5$.
			}
        \label{fig:h5FSS}
        %\end{center}
        }
    }
\end{figure}

\begin{figure}[ht]
    {
    \parbox[c]{\textwidth}
        {
 	\begin{subfigure}{0.46\textwidth}
         	\centering 
         	\includegraphics[angle=0, width=\textwidth]{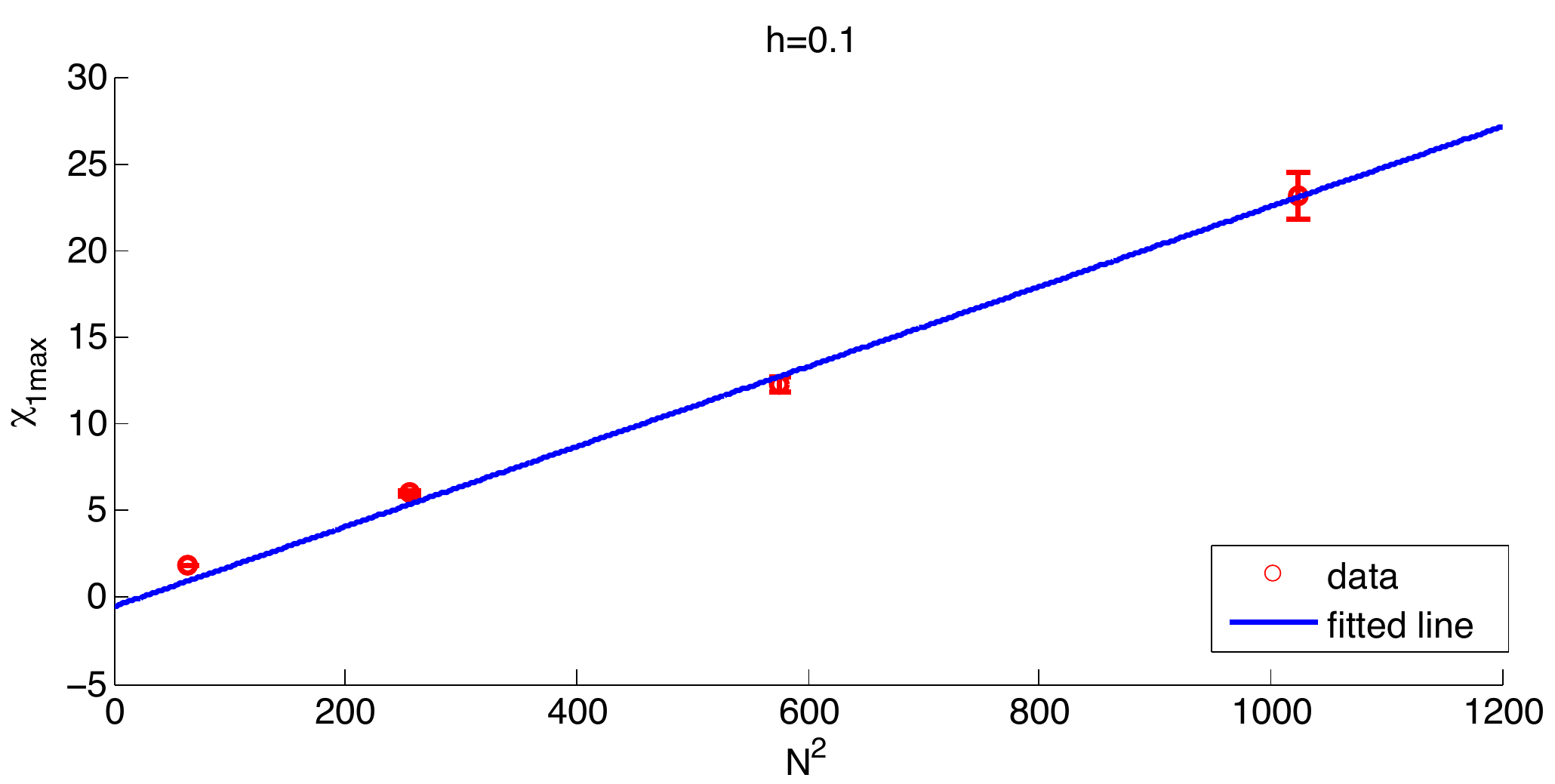}
 	\caption{  Maximum of the first component of susceptibility vs. $N^2$.}
 	\end{subfigure}
	\begin{subfigure}{0.45\textwidth}
		 \centering
        	\includegraphics[angle=0, width=\textwidth]{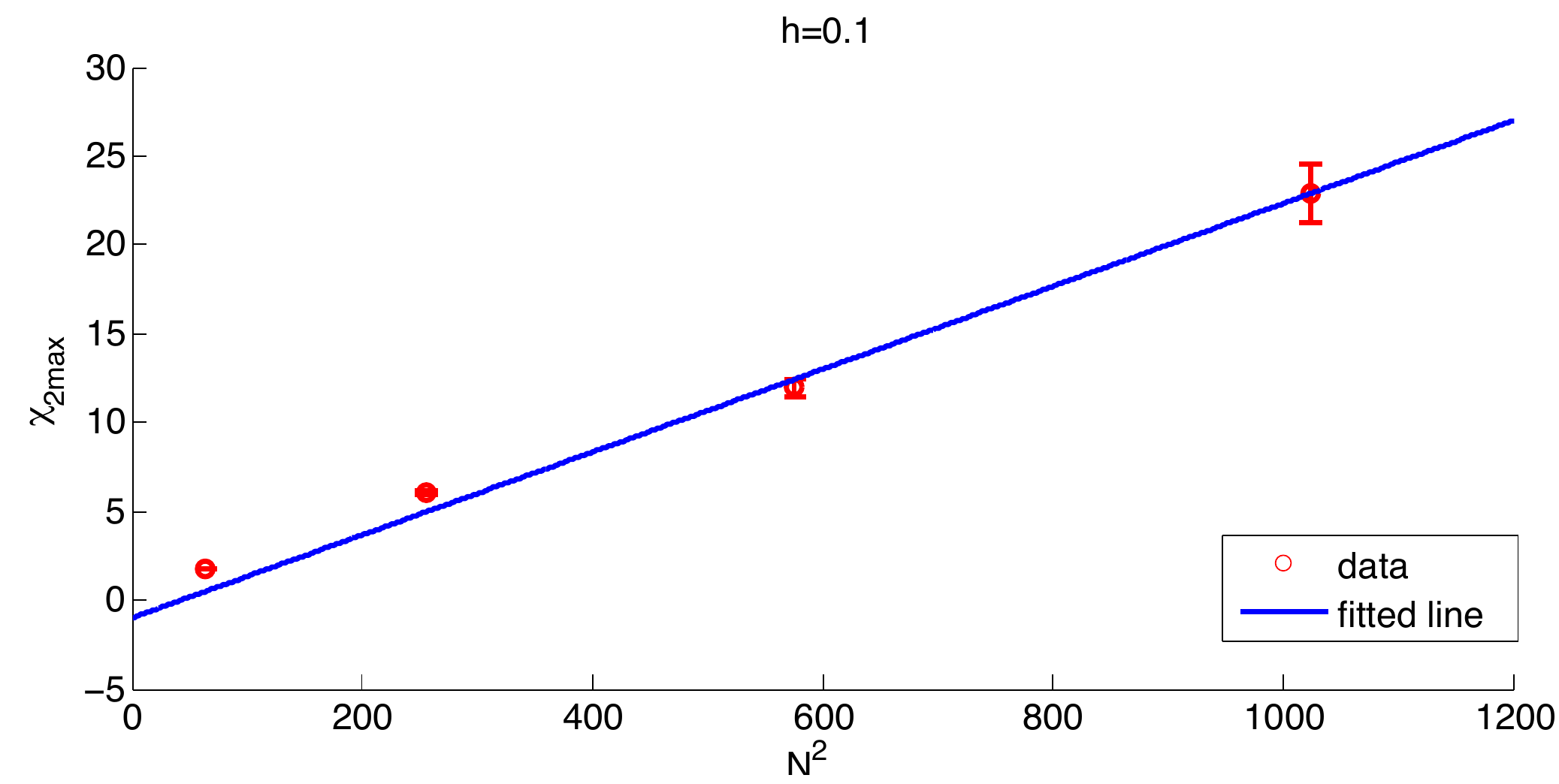}
		\caption{ Maximum of the second component of susceptibility vs. $N^2$.}
	\end{subfigure}
	\begin{subfigure}{0.46\textwidth}
        	\centering
        	\includegraphics[angle=0, width=\textwidth]{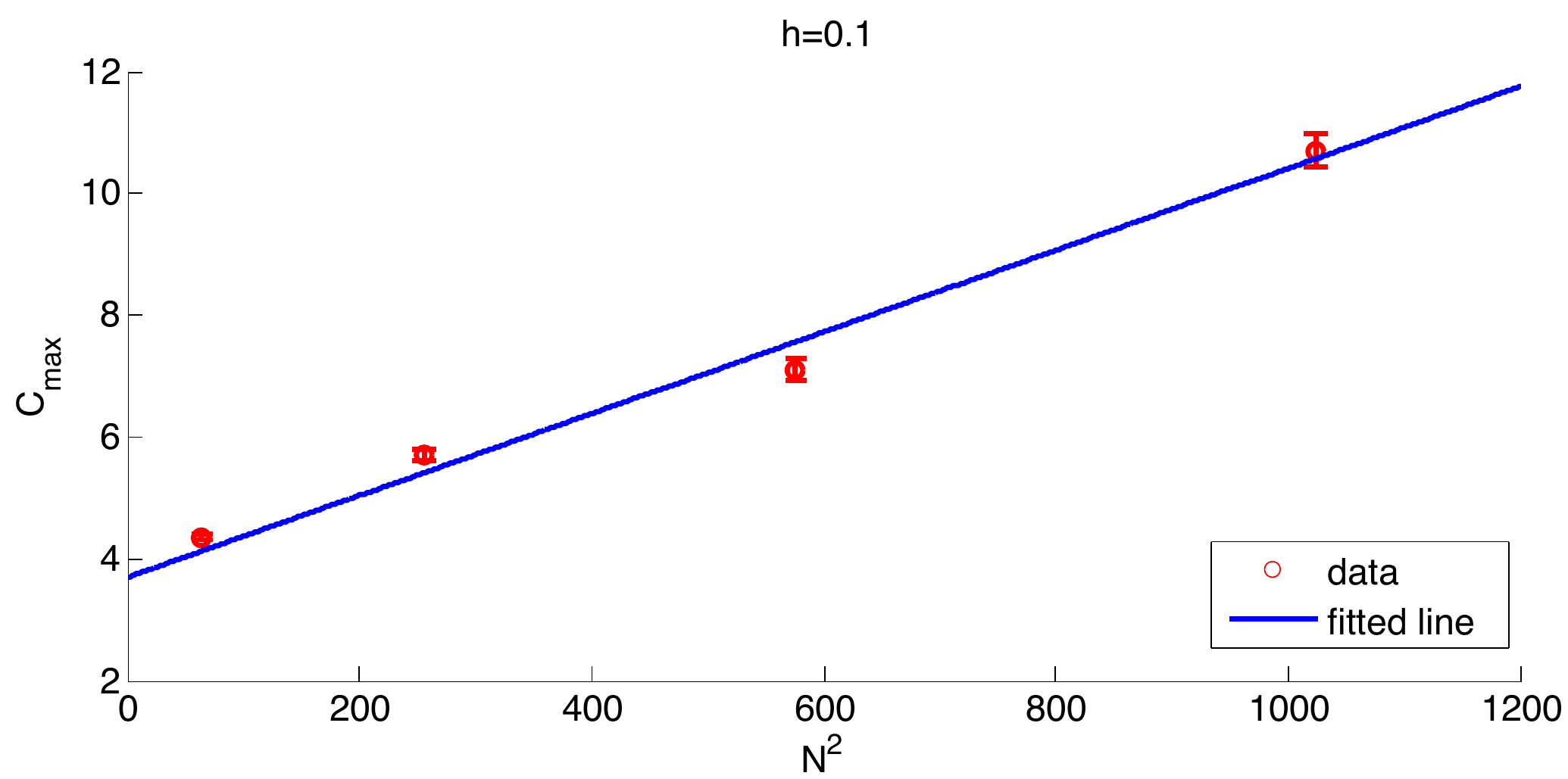}
		\caption{ Maximum of the specific heat per site vs. $N^2$.}
	\end{subfigure}
		\hfil
  \caption 
      {   Finite-size scaling of $\chi^i_{max}(N)$, $i=1,2$, and $C_{max}(N)$ for $h=0.1$.
			}
        \label{fig:h01FSS}
        %\end{center}
        }
    }
\end{figure}

On Figures~\ref{fig:h1FSS}, \ref{fig:h5FSS}, and \ref{fig:h01FSS}, we show, for $h=1$, $h=0.5$, and $h=0.1$, respectively, the scaling of the maxima, with respect to the temperature, of $\chi^1,\chi^2$, and $C$ with $N$ for $N=8,16, 24, 32$. The values of $C_{max}, \chi^{1}_{max}$ and $\chi^2_{max}$ are determined by fitting the raw data for each quantity with a degree 4 polynomial weighted with the error bars of the data (thus higher error bars have lower weights) and taking the maximum of that polynomial.  The error bars in $C_{max}, \chi^{1}_{max}$ and $\chi^2_{max}$  are the two-sigma (95\% confidence level) errors of the curve fitting.  Figures~\ref{fig:h1FSS}, \ref{fig:h5FSS}, and \ref{fig:h01FSS} indicate that the data for all $N$ and $h$ are fitted well with the linear dependence on the volume expected of a first order transition.

\begin{figure}[ht]
    {
    \parbox[c]{\textwidth}
        {
 	\begin{subfigure}{0.5\textwidth}
         	\centering 
         	\includegraphics[angle=0, width=\textwidth]{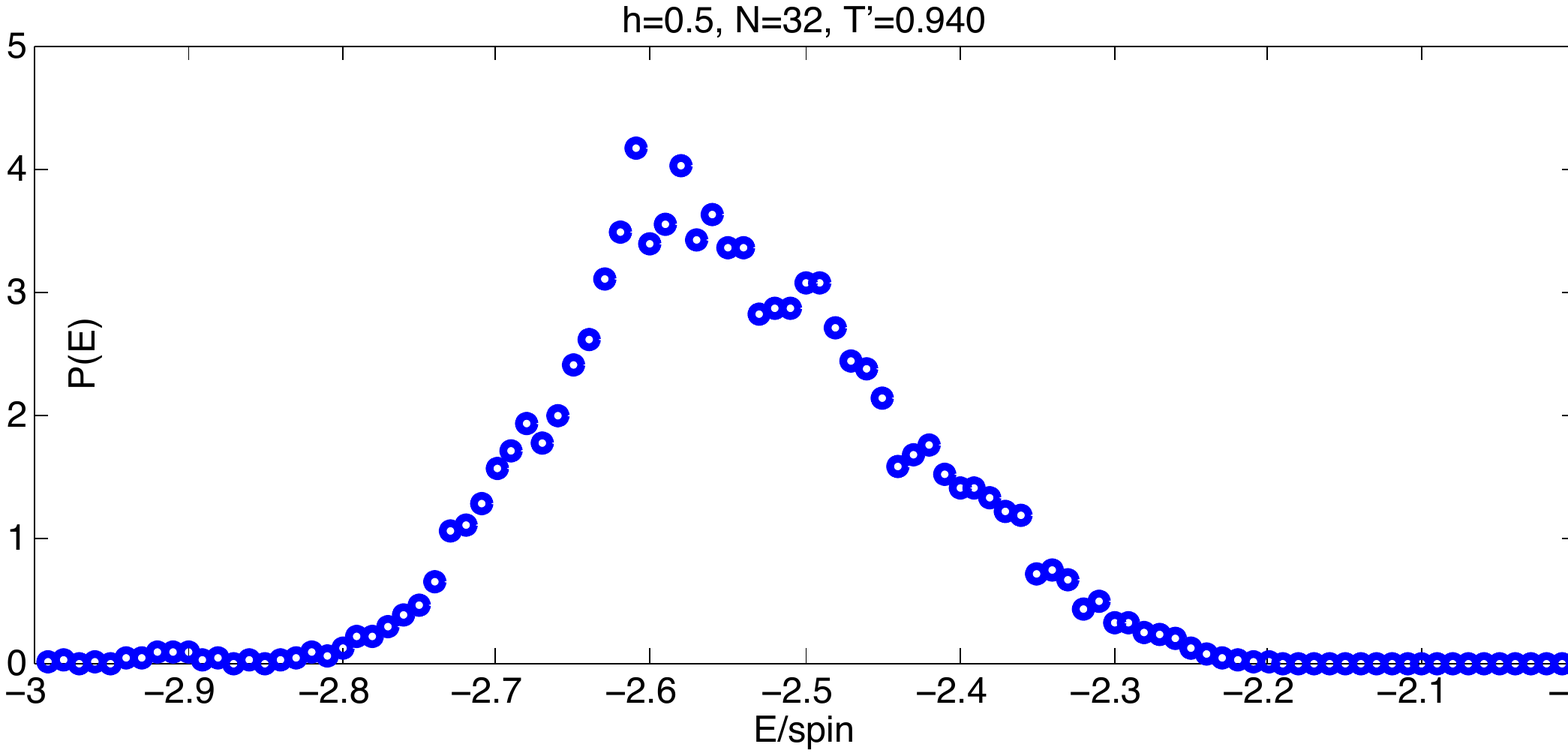}
 	\caption{Probability distribution of energy for $T'<T'_c$.}
 	\end{subfigure}
	\begin{subfigure}{0.5\textwidth}
		 \centering
        	\includegraphics[angle=0, width=\textwidth]{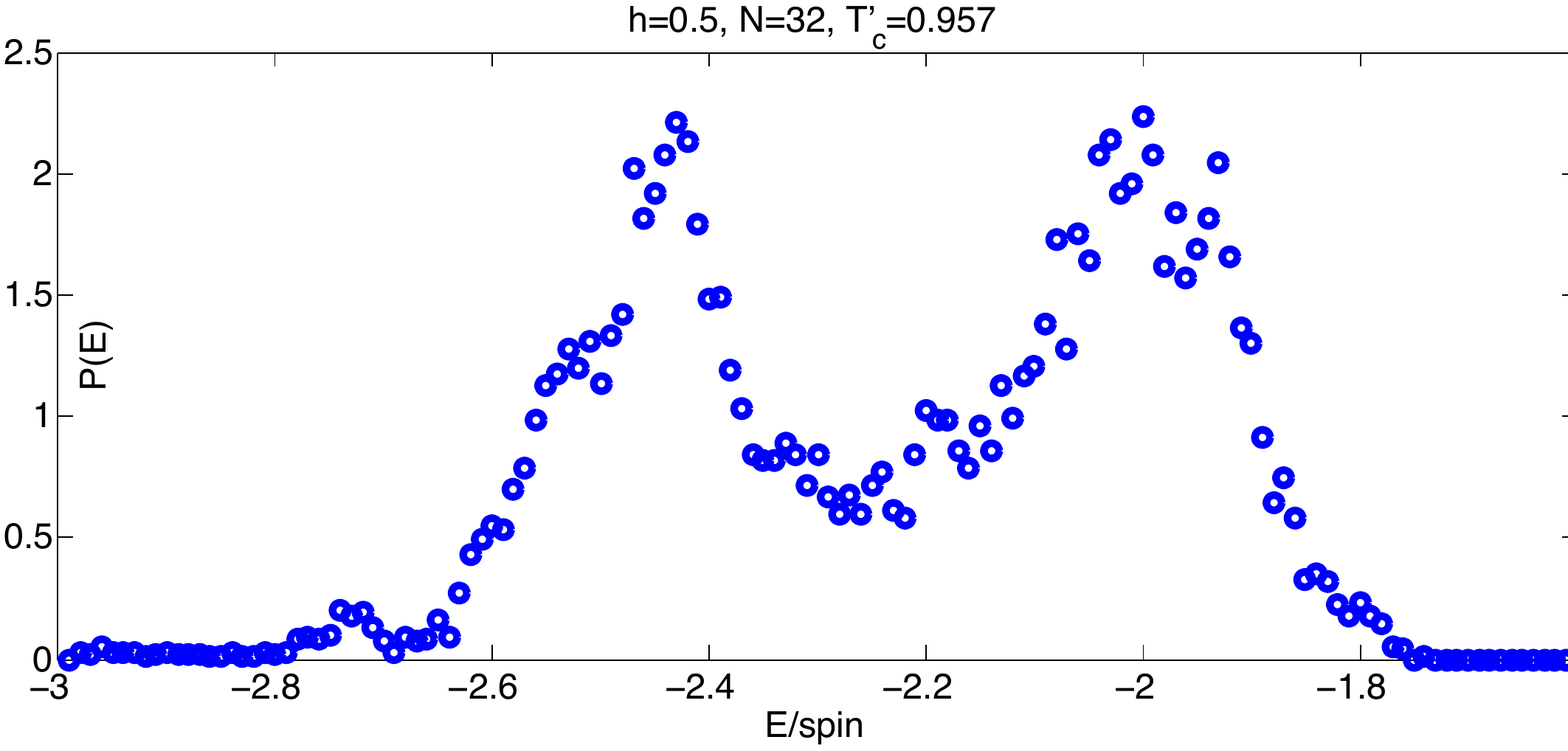}
		\caption{Probability distribution of energy for $T'=T'_c$.}
	\end{subfigure}
	\begin{subfigure}{0.5\textwidth}
        	\centering
        	\includegraphics[angle=0, width=\textwidth]{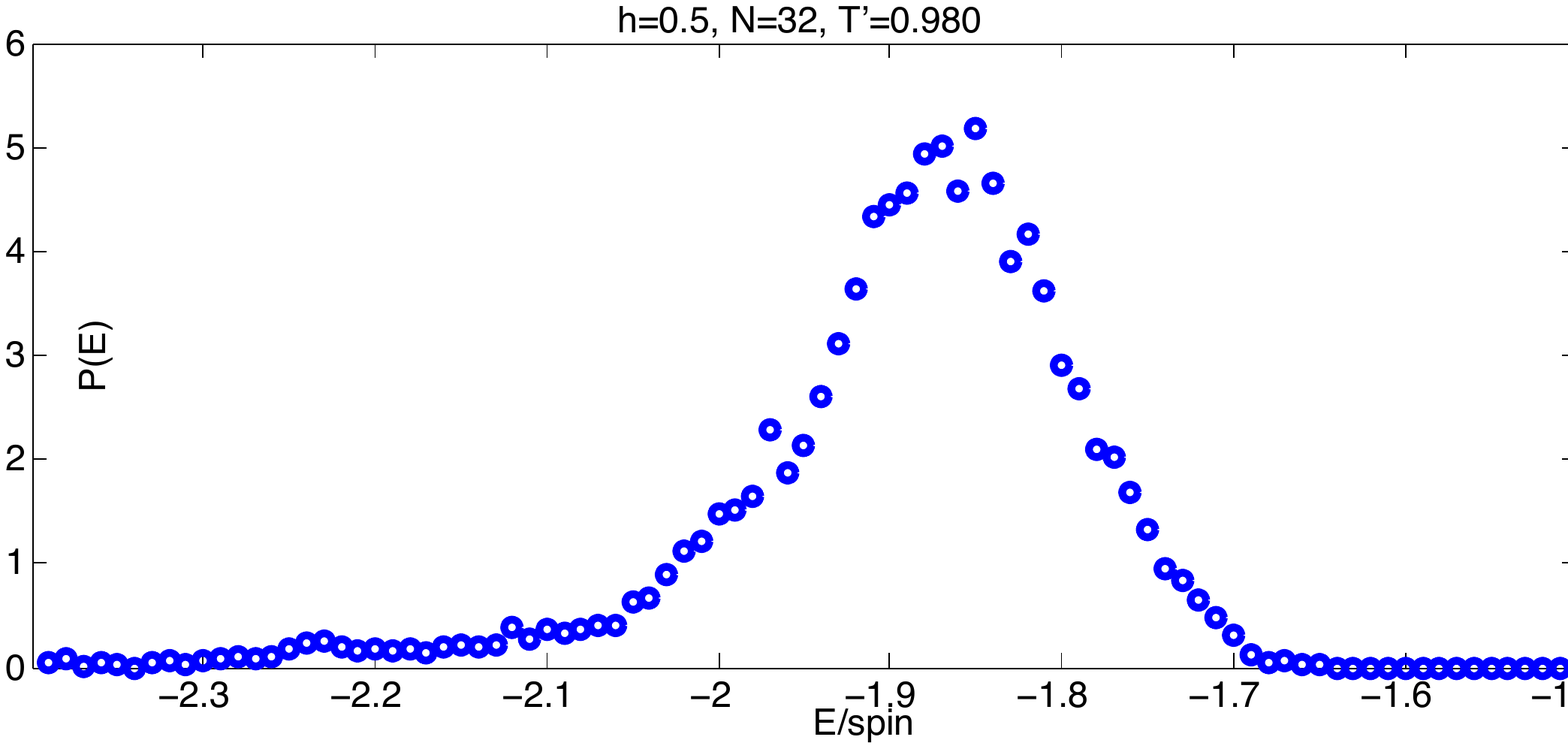}
		\caption{Probability distribution of energy for $T'>T'_c$.}
	\end{subfigure}
		\hfil
  \caption 
      {Probability distribution of the energy per site for $h=0.5$, $N=32$ for subcritical, critical and supercritical temperatures. The double peak distribution at $T'_c$ indicates phase coexistence and a first-order phase transition. The distance between the peaks is consistent with the latent heat calculated from the slope of the $C_{max}(N)$ vs. $N^2$ line, see the main text.
			}
        \label{fig:h5ET}
        %\end{center}
        }
    }
 \end{figure}

As a further check on the conjectured first order nature of the transition, we studied the probability distribution of the energy per spin. The histogram we plot was generated in a Monte Carlo run using $5700$ Monte Carlo sweeps (after waiting $4000$ sweeps for equilibration and then measuring the energy after every sweep) for $N=32$, $h=0.5$ for several values of $T'$, of which we show only three. Two distinct peaks are clearly visible in the energy probability distribution for $T'=T'_c$  on Figure~\ref{fig:h5ET}b. Conversely,   only one peak is present at either below or above $T'_c$, with its energy shifted in the right direction, see Figures~\ref{fig:h5ET}a and \ref{fig:h5ET}c. The order of magnitude of the distance between the peaks is consistent with the estimate of the latent heat obtained using the slope in (\ref{cmax}): taking $T'_c \sim 0.957$ for $h=0.5$, we find, using (\ref{cmax}) and Figure~\ref{fig:h5FSS}(c), $E_+ - E_- \simeq .38$. This value is consistent with the distance between the peaks, about $0.4$, estimated from Figure~\ref{fig:h5ET}, and provides a check on our data and the first-order transition hypothesis.

For $h=0.1$, on the other hand, we could not detect distinct peaks in the energy distribution and we do not show these results. This lack of distinct peaks could be because the $h=0.1$ phase transition is a very weak  first-order one, with small latent heat,\footnote{Indeed, the slope on Figure~\ref{fig:h01FSS}(c) is much smaller compared to the one for $h=1$ or $0.5$ with similar $T_c$. Assuming a first order transition and thus eq.~(\ref{cmax}), we can estimate the latent heat to be $\simeq 0.14$.} or because  for small $h$ the transition is continuous with $\gamma/\nu$ smaller but close to 2. The precision of our data does not allow us to be definite on this issue, which requires further work to settle.
 
\begin{figure}[ht]
          	\includegraphics[angle=0, width=\textwidth]{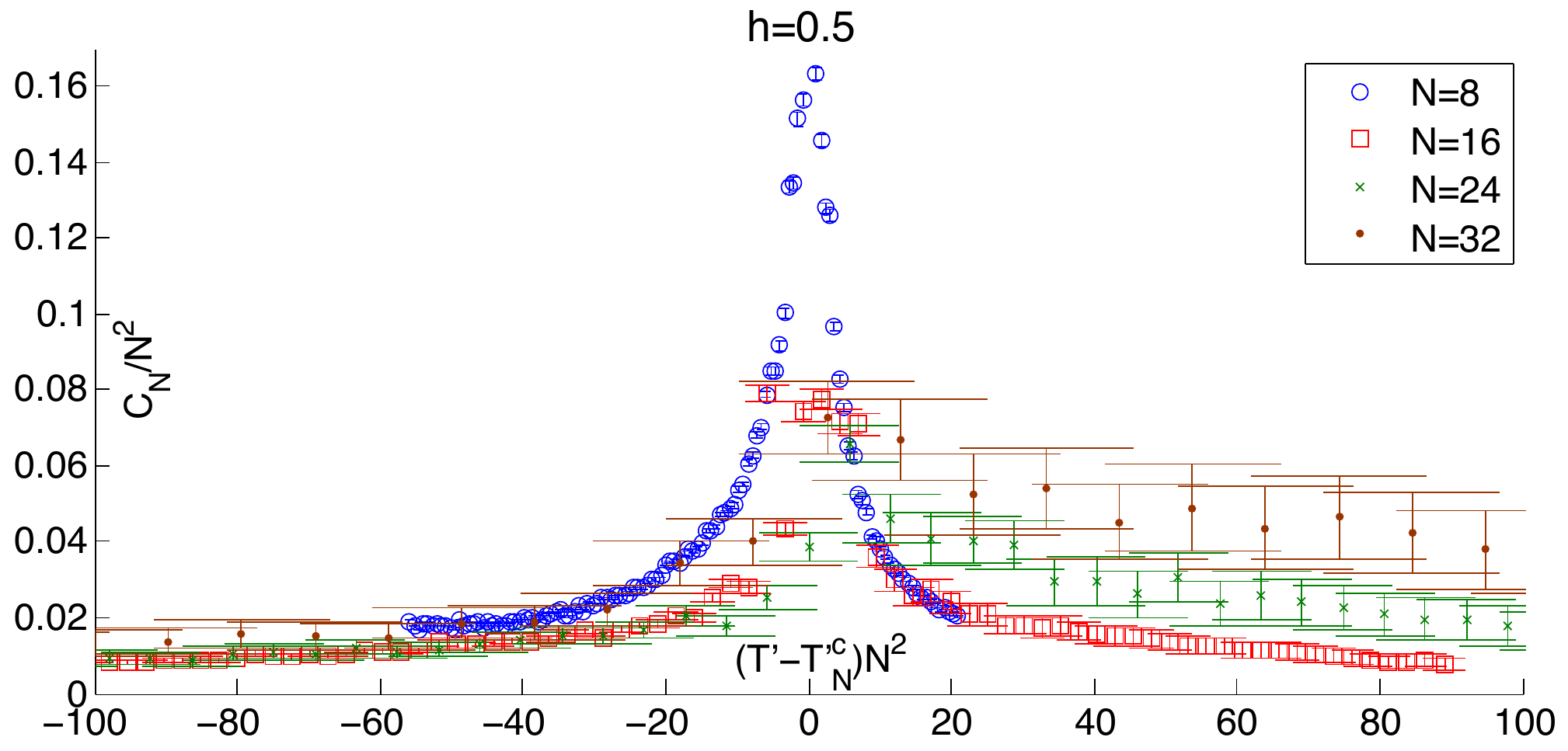}
		\caption{A plot of $C(N,T') N^{-2}$ vs. $(T' - T'_c(N)) N^2$. The data for different $N$ shows a reasonable collapse, except for the smallest  $N=8$ volume.}
	        \label{fig:data}
  \end{figure}

 Finally, we can test another prediction of the thermodynamic theory of finite-volume scaling in first order transitions. As Ref.~\cite{Challa:1986sk} showed,  for sufficiently large volumes and sufficiently close to $T'_c$, the heat capacity ${C(N,T') \over N^2}$ is expected to be a universal function of $(T'-T'_c(N)) N^2$---in other words, when plotted, the curves for different $N$ should collapse. To test this, we plot $C(N,T') N^{-2}$ as a function of $(T'-T'_c(N)) N^2$ for $N = 8,16, 24,32$, for $h=0.5$,  on Figure~\ref{fig:data}. The collapse of the  curves for different $N$ (apart from the smallest $N=8$) onto each other is reasonably good  for $T'<T'_c$, but both the statistics and the collapse are  poorer in the high-temperature phase.
  
  In summary, we have presented ample evidence that for moderately large values of the bion fugacity (for sure, $h\ge0.5$, but we have not simulated any $h$ between $0.1$ and $0.5$) the deconfinement transition is a first-order one. Larger values of the bion fugacities are expected when the decompactification limit $L \rightarrow \infty$ at fixed strong coupling scale $\Lambda$ is approached, as the bion action becomes smaller upon decompactification. Precisely a   first-order deconfinement transition was found in lattice simulations of the full QCD(adj)  $n_f=2$  gauge theory \cite{Karsch:1998qj}. 
  In the next Section we will summarize our findings, outline avenues for future studies, and speculate on the phase diagram in the $\beta$-$L$ plane for various values of $n_f$.

  \section{Conclusions and outlook}
\label{conclusions}
 
 \subsection{Summary of results}
 
This paper is devoted to a study of the phase transition in the ``affine" XY-spin model dual to the $SU(3)/\Z_3$ QCD(adj) theory on $\R^2 \times \S^1_\beta \times \S^1_L$, at small $L$, with a particular emphasis on the nature of the deconfining phase transition. 
 In our Monte Carlo study, we found that as  the temperature is increased, there is a single  deconfining and discrete-chiral restoration phase transition, which is discontinuous at least for $h \ge 0.5$, as found for the deconfinement transition in $SU(3)$ 4d QCD(adj) \cite{Karsch:1998qj} (with $n_f=4$). Finite-size scaling of the susceptibility and heat capacity and a study of the probability distribution for the energy were used to corroborate our conclusion.  
 
We  now continue with  comments on possible future work and speculations.

\subsection{Further studies of  (selfdual) electric-magnetic Coulomb gases}

The partition function of QCD(adj)  on $\R^2 \times \S^1_\beta \times \S^1_L$  in the 
regime (\ref{temperaturerange})  can be cast as   the grand canonical partition function of a multi-component Coulomb gas of electrically and magnetically charged particles interacting through logarithmic (dual) Coulomb and Aharonov-Bohm phase interactions. Schematically, it can be written as:
\begin{eqnarray}
\label{z1}
Z&=\sum\limits_{m,n}^{\infty}\frac{y^m\tilde y^n}{m!n!} \int \prod_{i,j} dx_i dy_j e^{-S_{bion}-S_{W}-S_{W/bion}}\,.
\end{eqnarray}
Here, $y$ and $\tilde{y}$ are the $W$-boson and bion fugacities and the sums/integrals are over   arbitrary numbers and spatial distribution of the various kinds of $W$-bosons and bions. The actions $S_{bion}$, $S_{W}$, and $S_{W/bion}$ contain their electric and magnetic Coulomb and Aharonov-Bohm interactions (an explicit expression for these can be found in Eq.~(4.9, 4.11, 4.13) in \cite{Anber:2011gn}). 

We show Eq.~(\ref{z1}) here only to note that the electric-magnetic Coulomb gas partition function (\ref{z1}) can be given a different description, which may be useful for analytical studies. To write it down, we introduce a two-compoenent, two-dimensional scalar field $\vec\Phi=\left(\Phi_1,\Phi_{2}\right)$ and its dual field  $\vec\Theta=\left(\Theta_1,\Theta_2\right)$. These obey the equal time commutation relation $[\Theta^i(x), \Phi^j(y)] = - i \delta^{ij} \theta(x-y)$, where $\theta$ is the Heaviside theta function, so that
$\vec\Pi_{\vec\Phi} = \partial_x \vec\Theta$ is the momentum conjugate to $\vec\Phi$; see, e.g., \cite{Tsvelik}. Then, the 
 2d Coulomb gas partition function (\ref{z1}) can be represented as the Euclidean vacuum functional 
of a one-dimensional quantum system with Hamiltonian density:
\begin{eqnarray}
\label{v1dual}
{\cal H}=\frac{1}{2}  (\partial_x\vec \Phi)^2+\frac{1}{2}  (\partial_x\vec{\Theta})^2-\sum\limits_{i=1}^{N_c=3}\left[\tilde y\cos\left[\frac{4\pi\sqrt{LT}}{g}(\vec\alpha_i-\vec\alpha_{i-1})\vec\Phi\right]+ y \cos\left[\frac{g }{\sqrt{LT}}\vec\alpha_i\vec\Theta\right]\right]. 
\end{eqnarray}
Here, $\tilde y$ and $ y$ are the magnetic and electric fugacities also appearing in (\ref{z1}), $g\equiv g_4(L)$, $L$, and $T$ are the  gauge theory quantities appearing in (\ref{kappasu3},\ref{bionfugacity}).   The deconfinement phase transition in the $SU(3)/\Z_3$ QCD(adj) theory is the quantum phase transition  in the theory with Hamiltonian density (\ref{v1dual}). As already advertised, the theory  exhibits electric-magnetic duality, which, in the  variables of (\ref{v1dual}), takes the simple form:\footnote{Invariance of (\ref{v1dual})  under (\ref{h1dual}) follows after using the expressions for the $SU(3)$ affine roots $\vec\alpha_i$ given in Footnote \ref{weights}. Note that a Hamiltonian identical to (\ref{v1dual}) describes also the Coulomb gas for $SU(N_c)$ QCD(adj), the only difference being that there are $N_c -1$ fields ($\vec{\Phi}, \vec{\Theta}$) and that the sum over the affine roots now extends from $1$ to $N_c$. It is easily seen  that for $N_c > 3$ there is no analogue of the electric-magnetic duality (\ref{v1dual}); the physical reasons are explained in \cite{Anber:2011gn}. One consequence of this lack of self-duality is that  for $N_c > 3$ the couplings ($\sim g$ and $\sim 1/g$) inside the two cosine-potentials in (\ref{v1dual}) renormalize differently.  }
\begin{eqnarray}
\Phi_2  \leftrightarrow  \Theta_1, ~~~~  \Phi_1 \leftrightarrow \Theta_2, ~~~~
 \tilde{y}  \leftrightarrow  y, ~~~~ {g  \over\sqrt{L T}} \leftrightarrow   {4 \pi \sqrt{3 L T}  \over g }.  \label{h1dual}
\end{eqnarray}
The last relation in (\ref{h1dual}) is the already mentioned $\kappa \leftrightarrow {12\over \kappa}$ symmetry, with $\kappa$ given in (\ref{kappasu3}), and shows that this is also a high-$T$/low-$T$ duality. At the self-dual point, it is easy to see that both the electric and magnetic  potential couplings (i.e., the two cosines) in (\ref{h1dual}) are relevant (in stark contrast to the $SU(2)$ QCD(adj) case \cite{Anber:2011gn}, where they are marginal).

Now that we have presented the selfdual sin-Gordon formulation (\ref{v1dual}) of the model, it is of interest to contrast the Monte Carlo results on the phases of the ``affine" XY-spin model with previous analytic studies of   electric-magnetic Coulomb gases in 2d. The leading order, in bion and $W$-boson fugacities, renormalization group equations (RGEs) of the Coulomb gas dual to  $SU(3)$ QCD(adj) \cite{Anber:2011gn} have a fixed point for the electric (and magnetic) coupling $\kappa$. The fixed point occurs at the self dual point  $\kappa_* = \sqrt{12} \simeq 3.46$. However, this analysis was (rightly) not considered conclusive---at  $\kappa_*$, the bion and $W$-bion fugacities are both relevant, indicating that a higher-order (indeed, strong coupling) analysis is necessary. Our present study indicates that no fixed point is present, at least for sufficiently high fugacities; instead, one finds a first-order phase transition. 
These findings would also be in contrast with conclusions drawn by a straightforward extension of an earlier analysis of RGEs to the next nontrivial order in fugacities. In Ref.~\cite{Boyanovsky:1990iw},  a similar self-dual Coulomb gas associated with the $SU(3)$ algebra was studied, and the RG  analysis was performed using the selfdual sin-Gordon representation similar to (\ref{v1dual}). The model studied there is not identical to our: as opposed to (\ref{v1dual}), in \cite{Boyanovsky:1990iw}  only $\vec\alpha_i$'s appear in the first cosine, rather than $\vec\alpha_i - \vec\alpha_{i-1}$; accounting for this difference is trivial and allows us to recast the results for the RGEs of \cite{Boyanovsky:1990iw} to our model. Taking them at face value, one might be tempted to conclude that there is a fixed point; however, the approximations used in \cite{Boyanovsky:1990iw} to derive the RGEs also do not strictly hold for  (\ref{v1dual}). The lack of fixed point that we find  illustrates the danger of relying on an uncontrolled approximation to infer the existence of  a fixed point.\footnote{As already mentioned, the apparent lack of a fixed point in $SU(3)$ QCD(adj) is in  contrast with the   $SU(2)$  case, where a continuous transition was found. A line of self-dual fixed points  extends to weak coupling (small fugacities), thus allowing a controlled calculation of various critical exponents \cite{Anber:2011gn} (in fact,   using 2d CFT methods, one can show that the phase transition in the spin model dual to  $SU(2)$(adj) is second order for all values of the fugacities, see the first reference in \cite{Lecheminant:2002va}). The spin model dual to $SU(2)$(adj) case is a single-component XY-model with a $\Z_4$-preserving perturbation. Continuously varying critical exponents, corresponding to the fixed-line behavior, have  also been seen in lattice  simulations of the  $\Z_4$-model \cite{Z4paper}.} 

In the future, it would be of interest to further study (\ref{v1dual}), both for $N_c = 3$ and higher, with an aim towards understanding  the small-$h$ behavior (e.g., continuous vs. discontinuous transition) that we were inconclusive about. This could possibly  be done by  applying 2d CFT methods, as in  \cite{Lecheminant:2002va},  to   (\ref{h1dual})  and  is  an interesting task for future work.
 
There are also some obvious generalizations of the present Monte Carlo study. Notably,  
the $SU(N_c>3)$ QCD(adj) theories on $\R^2 \times \S^1_\beta \times \S^1_L$ are also dual to electric-magnetic Coulomb gases, given by the higher-$N_c$ generalizations of (\ref{v1dual}). Due to the ``composite" nature of magnetic bions, these gases do not exhibit electric-magnetic duality for $N_c>3$ and, at present,  we have no clues as of the nature of the deconfinement transition in these theories (the leading-order RGEs do not give any useful information and there is no self-dual point). It should be relatively straightforward, if more resource consuming, to pursue this question numerically. 

\subsection{The global $\beta$-$L$ phase diagram}
\label{phasediagram}

\begin{figure}[ht]
    {
    \parbox[c]{\textwidth}
        {
	\centering
%	\begin{subfigure}{0.55\textwidth}
  %      	\includegraphics[angle=0, width=\textwidth]{h030000M1.pdf}
%		\caption{$h=0$}
%	\end{subfigure}
	\begin{subfigure}{0.47\textwidth}
	\centering
        	\includegraphics[angle=0, width=\textwidth]{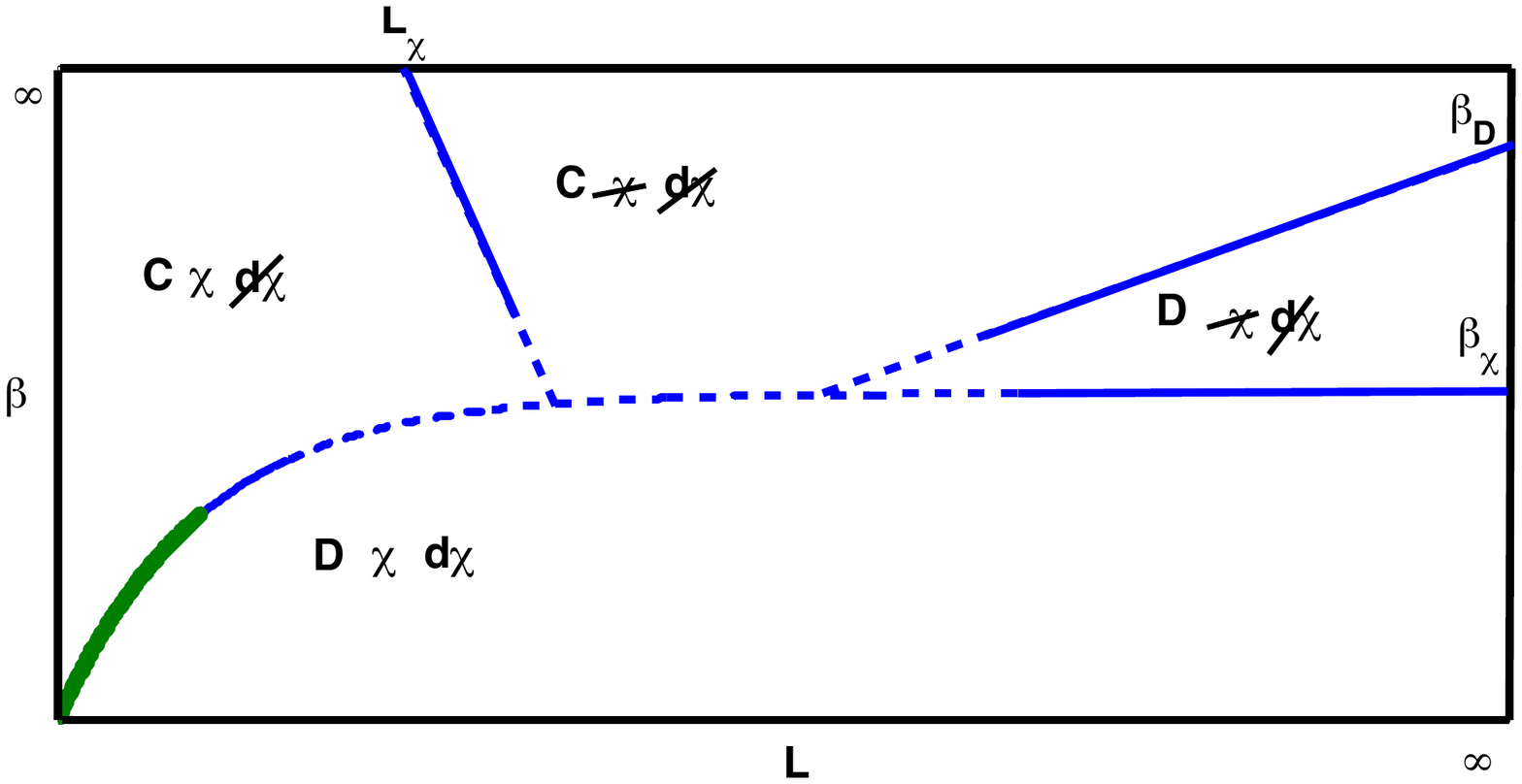}
		\caption{One simplest possible $\beta$-$L$ phase diagram for $n_f < n_f^{\it cr.}$.  }
	\end{subfigure}
 \centering
	\begin{subfigure}{0.47\textwidth}
	\centering
        	\includegraphics[angle=0, width=\textwidth]{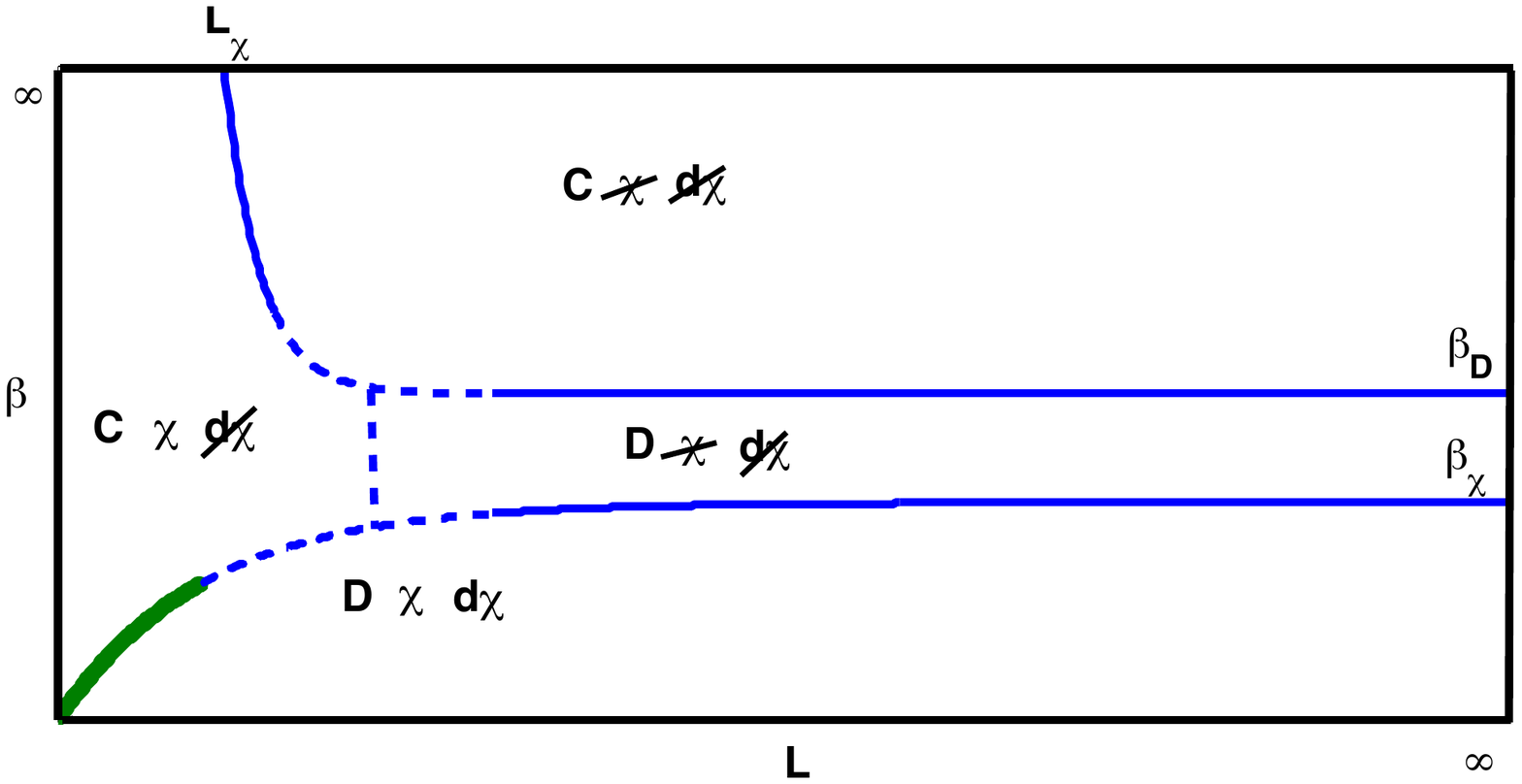}
		\caption{The other simplest possible $\beta$-$L$ phase diagram for $n_f < n_f^{\it cr.}$.}
	\end{subfigure}
		\hfil
	\caption{\small The two simplest possible phase diagrams connecting the small-$L$ and large-$L$ behavior for theories which are confining in the infinite-$L$ zero-$T$ (infinite-$\beta$ limit). The thick red lines in the lower left hand corner denote the location of the discontinuous   phase transition studied here.
  In both diagrams, here and in Figure 2, ``C" and ``D" refer to confined and deconfined phases, while ``$\chi$" and ``$d\chi$" denote the realization of the $SU(n_f)$ continuous and $Z_{2 N_c n_f}$ discrete chiral symmetries in the various regions. $\beta_D$ and $\beta_\chi$ are the deconfinement  and $SU(n_f)$-restoration temperatures, respectively, in the infinite-$L$ theory. $L_\chi$ denotes a critical radius $L$ beyond which $SU(n_f)$ is broken (at zero temperature).
}
        \label{fig:phase}
       % \end{center}
        }
      }
\end{figure}

\begin{figure}[ht]
    {
    \parbox[c]{\textwidth}
        {
	\centering
%	\begin{subfigure}{0.55\textwidth}
  %      	\includegraphics[angle=0, width=\textwidth]{h030000M1.pdf}
%		\caption{$h=0$}
%	\end{subfigure}
		 \centering
	\begin{subfigure}{0.5\textwidth}
	\centering
        	\includegraphics[angle=0, width=\textwidth]{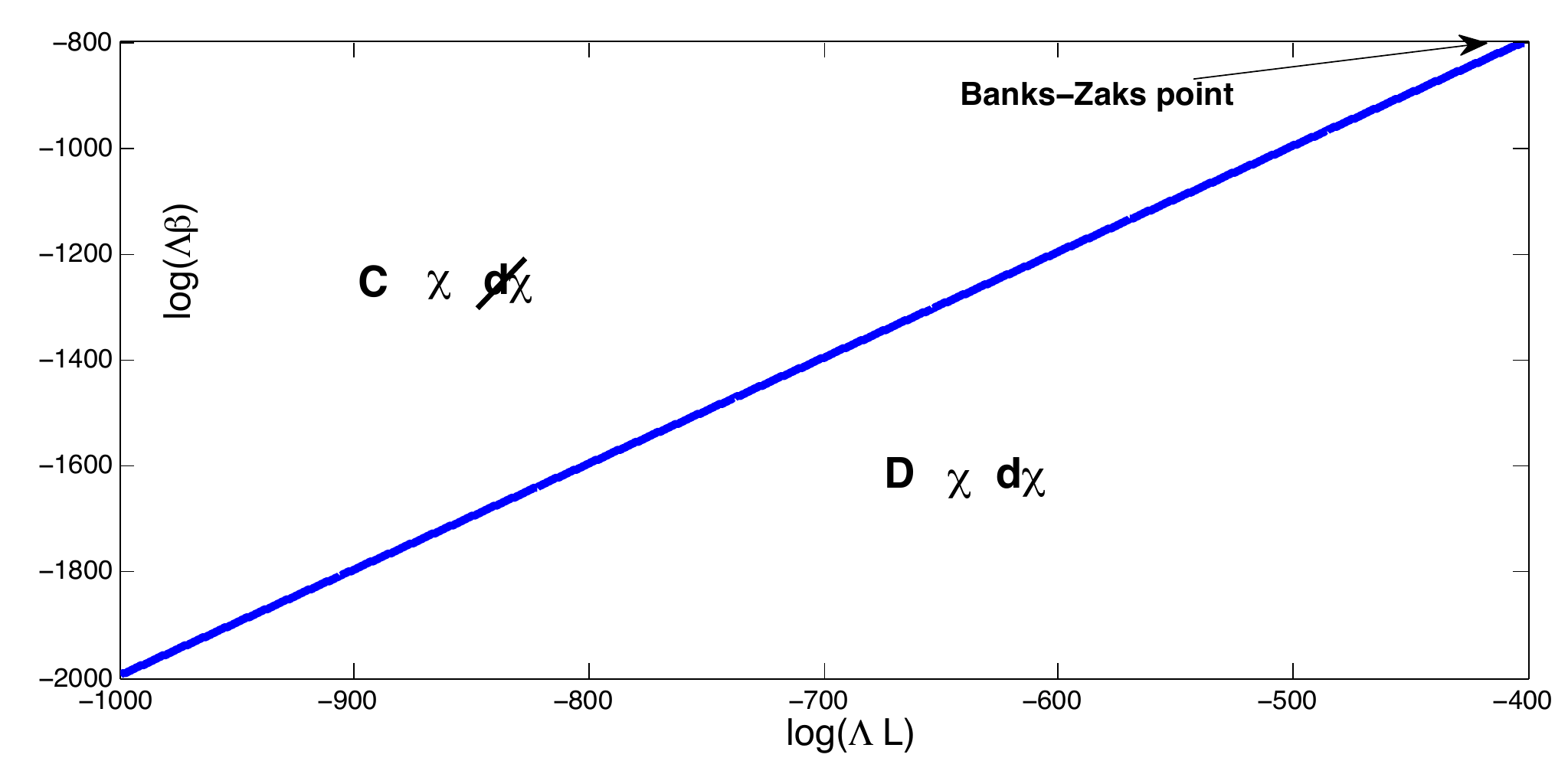}
			\end{subfigure}
	\hfil
	\caption{\small
  The $n_f=5$ theory is (most likely) conformal with a, presumably, sufficiently weak-coupling Banks-Zaks like fixed point $\alpha_*$. If so, upon increasing $L$, the semiclassical analysis is always applicable. In particular, it shows that the conformal zero-temperature theory confines at any finite $L$ with an exponentially small mass gap \cite{Poppitz:2009uq}. There is a thermal deconfinement and $d\chi$-breaking transition with critical temperature given, for $SU(2)$, by $T_c(L) ={ \alpha(L) \over 2 L}$ \cite{Anber:2011gn}. The transition is second order for $SU(2)$. The line drawn on the figure corresponds to $T_c(L)$ drawn with the two-loop $SU(2)$ coupling.
 The transition for $SU(3)$ is---probably, see Section 2.3---discontinuous  and occurs at temperatures  $T' \sim 1$, parametrically giving $T_c(L) \sim {g_4^2(L)\over L}$, similar to the $SU(2)$ case. The semiclassical theory, now expected to be valid at all $L$, predicts that the discrete chiral symmetry, but not $SU(n_f)$, also breaks at the deconfinement transition.}
        \label{fig:phase5}
       % \end{center}
        }
      }
\end{figure}

We end with a discussion of the hardest and most interesting question: the relation of the small-$L$ deconfinement transition to that in the large-$L$ theory. We first recall that  in $SU(3)$ QCD(adj),  ``large-$L$" lattice studies  have found that the deconfinement and continuous nonabelian and discrete   chiral symmetry transitions do not coincide and that deconfinement is a first order transition \cite{Karsch:1998qj}.\footnote{Thus, the order of the small-$L$ deconfinement transition for QCD(adj) is the same as  the 4d theory transition. This is in contrast to deformed pure-Yang-Mills theory, where, in $SU(N_c)$ theories it was found that the small-$L$ transition is second order, in the universality class of $Z_{N_c}$ parafermion 2d CFT, see \cite{Lecheminant:2002va}.} 
Upon increasing the temperature, the deconfinement transition happens first, while chiral symmetry is restored only at higher temperatures. Based on this knowledge as well as our small-$L$ studies, we have drawn,  on Figure~\ref{fig:phase}(a) and (b),
two possible phase diagrams of $SU(3)$ QCD(adj)---for $n_f$ such that the infinite-$L$ theory is confining. Zero-temperature confinement occurs for $n_f < n_f^{\it cr.}$  (here $n_f^{\it cr.}$ is the critical number of Weyl adjoints above which the theory is conformal, presumably $\sim 4$ or $5$ for QCD(adj)). For $n_f = 5$,  most likely a conformal case, the simplest possible phase diagram is given in Figure~\ref{fig:phase5}  and the arguments leading to it are  discussed further below.

The phase diagrams on Figure~\ref{fig:phase}  were drawn using the  information that we have on the phases from large-$L$ lattice studies and from our small-$L$ semiclassical and numerical studies. For large-$L$, our phase diagrams assume distinct deconfinement and chiral transitions as in \cite{Karsch:1998qj} and in drawing the phase diagrams we have assumed that only three phases can coexist (barring accidents), since for exactly massless fermions there are only two parameters to vary, $\beta$ and $L$. As already mentioned, the small-$L$ dynamics of QCD(adj) always preserves the $SU(n_f)$ chiral symmetry. In the confining theories, the continuous (as well as discrete) chiral symmetry is  believed to be broken in the large-$L$ limit. Hence, an extension of the present methods to incorporate it is required---if one is to continue the small-$L$ studies to larger-$L$ ---for  generic $n_f$, see \cite{Poppitz:2009uq} for discussion.

There is one exception worth mentioning, however. For $SU(N_c)$ with $n_f=5$ adjoint Weyl fermions,  the theory is very likely conformal in the infinite-$L$ limit due to the existence of a Banks-Zaks infrared fixed point at (reasonably---i.e., numerically, but not parametrically) weak coupling. In this case, the phase diagram considerably simplifies, as shown in 
Figure~\ref{fig:phase5}. For $SU(2)$, the confined (and discrete chiral-broken)\footnote{We note that in the weak-coupling semiclassical regime, the order parameter for the ``$d \chi$" $Z_{2 N_c n_f}$ symmetry are the exponentials of the dual photons \cite{Unsal:2007vu}, $e^{i \vec{\alpha} \cdot \vec\sigma}$, while at strong coupling, local $SU(n_f)$-singlet but ``$d \chi$" nonsinglet order parameters can be built from multifermion operators like $\det\limits_{i,j} \lambda^{a \alpha}_j \lambda^a_{\beta j}$, where $a=1\ldots N_c^2 -1$, $\alpha, \beta=1,2$, and $i,j=1,..,n_f$ are adjoint, $SL(2,C)$-Lorentz, and flavor indices, respectively.} and deconfined (and discrete chiral-symmetric) phases are separated by the line $T_c(L)=\frac{\alpha(L)}{2L}$, where $\alpha={g^2\over 4\pi}$, up to exponentially small corrections, as was shown in \cite{Anber:2011gn}. 
The coupling $\alpha(L)$ is, to two-loop order:
\begin{equation}\label{twoloop}
\alpha(L)=\frac{4\pi}{\beta_0}\left[\frac{1}{\log(1/\Lambda^2L^2)} -\frac{\beta_1}{\beta_0^2}\frac{\log\log(1/\Lambda^2L^2)}{\left(\log(1/\Lambda^2L^2)\right)^2}\right]\,,
\end{equation}
 where $\Lambda$ is the strong coupling scale (in an IR conformal theory, this is the scale where the behavior of the running coupling switches from asymptotically free to conformal), $\beta_0=(22-4n_f)/3$ and   $\beta_1=(136-64n_f)/3$.  Eq.~(\ref{twoloop}), with  $n_f =5$, was used to draw the phase transition line $\beta_c(L) = {2L \over \alpha(L)}$ on Figure~\ref{fig:phase5}. We emphasize that the usual strong coupling problem mentioned in Section \ref{digression} is avoided because the coupling (\ref{twoloop}) runs, at large-$L$, into the fixed point at $\alpha_* = - {4 \pi \beta_0 \over \beta_1} = {\pi \over 23}  \simeq 0.136 $ rather than into the infrared ``Landau pole" (the bion action at the fixed point is ${2 \pi\over \alpha_*} \simeq 46$, and the semiclassical approximation is still expected to hold \cite{Poppitz:2009uq}).
 
The global---i.e., valid at any $L$---phase diagram for QCD(adj) for $n_f=5$ on Figure~\ref{fig:phase5} can be viewed as a prediction of our semiclassical studies and could, in principle, be verified by lattice simulations.
We end with the remark that,  one, not surprising, conclusion is that the phase diagram of QCD(adj)  is rather complicated and that, in this paper, we have only explored a small corner tractable by a combination of semiclassical and numerical methods.

\acknowledgments

We thank Marat Mufteev for collaboration at the early stages of this project and Mithat \" Unsal for comments on the manuscript. We acknowledge 
support by an NSERC Discovery Grant and by  an NSERC Undergraduate Student Research Award (for S.C.).
 Part of this work was made possible by the facilities of the Shared Hierarchical Academic Research Computing Network (SHARCNET:www.sharcnet.ca) and Compute/Calcul Canada.

\begin{thebibliography}{1}
 
   %\cite{Anber:2011gn}
\bibitem{Anber:2011gn} 
  M.~M.~Anber, E.~Poppitz and M.~\" Unsal,
``2d affine XY-spin model/4d gauge theory duality and deconfinement,''
  JHEP {\bf 1204}, 040 (2012)
  [arXiv:1112.6389 [hep-th]].
  %%CITATION = ARXIV:1112.6389;%%
 
 \bibitem{Seiberg:1996nz} 
  N.~Seiberg and E.~Witten,
   ``Gauge dynamics and compactification to three-dimensions,''
  In *Saclay 1996, The mathematical beauty of physics* 333-366
  [hep-th/9607163];
  %%CITATION = HEP-TH/9607163;%%
  
  %\cite{Aharony:1997bx}
%\bibitem{Aharony:1997bx} 
  O.~Aharony, A.~Hanany, K.~A.~Intriligator, N.~Seiberg and M.~J.~Strassler,
  ``Aspects of N=2 supersymmetric gauge theories in three-dimensions,''
  Nucl.\ Phys.\ B {\bf 499}, 67 (1997)
  [hep-th/9703110];%\cite{Davies:1999uw}
%\bibitem{Davies:1999uw} 
  
  N.~M.~Davies, T.~J.~Hollowood, V.~V.~Khoze and M.~P.~Mattis,
   ``Gluino condensate and magnetic monopoles in supersymmetric gluodynamics,''
  Nucl.\ Phys.\ B {\bf 559}, 123 (1999)
  [hep-th/9905015];
  %%CITATION = HEP-TH/9905015;%%
 %\cite{Davies:2000nw}
%\bibitem{Davies:2000nw} 
  
  N.~M.~Davies, T.~J.~Hollowood and V.~V.~Khoze,
   ``Monopoles, affine algebras and the gluino condensate,''
  J.\ Math.\ Phys.\  {\bf 44}, 3640 (2003)
  [hep-th/0006011].
  %%CITATION = HEP-TH/0006011;%%
 
 %\cite{Unsal:2007vu} 
\bibitem{Unsal:2007vu} 
  M.~\" Unsal,
 ``Abelian duality, confinement, and chiral symmetry breaking in QCD(adj),''
  Phys.\ Rev.\ Lett.\  {\bf 100}, 032005 (2008)
  [arXiv:0708.1772 [hep-th]];
  %%CITATION = ARXIV:0708.1772;%%
 %\cite{Unsal:2007jx}

%\bibitem{Unsal:2007jx} 
  M.~\" Unsal,
  ``Magnetic bion condensation: A New mechanism of confinement and mass gap in four dimensions,''
  Phys.\ Rev.\ D {\bf 80}, 065001 (2009)
  [arXiv:0709.3269 [hep-th]].
  %%CITATION = ARXIV:0709.3269;%%


  
  %%CITATION = ARXIV:0905.0634;%%
%\cite{Polyakov:1976fu}
\bibitem{Polyakov:1976fu} 
  A.~M.~Polyakov,
   ``Quark Confinement and Topology of Gauge Groups,''
  Nucl.\ Phys.\ B {\bf 120}, 429 (1977).
  %%CITATION = NUPHA,B120,429;%%
%\cite{Shifman:2008cx}\cite{Poppitz:2009kz}\cite{Poppitz:2009uq}\cite{Poppitz:2009tw}

\bibitem{Shifman:2008cx} 
  M.~Shifman and M.~\" Unsal,
   ``On Yang-Mills Theories with Chiral Matter at Strong Coupling,''
  Phys.\ Rev.\ D {\bf 79}, 105010 (2009)
  [arXiv:0808.2485 [hep-th]];
  %%CITATION = ARXIV:0808.2485;%%
  %\cite{Poppitz:2009kz}
 
 %\bibitem{Poppitz:2009kz} 
  E.~Poppitz and M.~\" Unsal,
 ``Chiral gauge dynamics and dynamical supersymmetry breaking,''
  JHEP {\bf 0907}, 060 (2009)
  [arXiv:0905.0634 [hep-th]];  
  %\cite{Poppitz:2009uq}


%\bibitem{Poppitz:2009tw} 
  E.~Poppitz and M.~\" Unsal,
  ``Conformality or confinement (II): One-flavor CFTs and mixed-representation QCD,''
  JHEP {\bf 0912}, 011 (2009)
  [arXiv:0910.1245 [hep-th]].
  %%CITATION = ARXIV:0910.1245;%%

%\cite{Poppitz:2009uq}
 \bibitem{Poppitz:2009uq} 
  E.~Poppitz and M.~\" Unsal,
  ``Conformality or confinement: (IR)relevance of topological excitations,''
  JHEP {\bf 0909}, 050 (2009)
  [arXiv:0906.5156 [hep-th]];
  %%CITATION = ARXIV:0906.5156;%%
  %\cite{Poppitz:2009tw}
  


%\cite{Dunne:2000vp}
\bibitem{Dunne:2000vp} 
  G.~V.~Dunne, I.~I.~Kogan, A.~Kovner and B.~Tekin,
   ``Deconfining phase transition in (2+1)-dimensions: The Georgi-Glashow model,''
  JHEP {\bf 0101}, 032 (2001)
  [hep-th/0010201];
   %\cite{Simic:2010sv}

%\bibitem{Simic:2010sv} 
  D.~Simi\' c and M.~\" Unsal,
  ``Deconfinement in Yang-Mills theory through toroidal compactification with deformation,''
  Phys.\ Rev.\ D {\bf 85}, 105027 (2012)
  [arXiv:1010.5515 [hep-th]].
  %%CITATION = ARXIV:1010.5515;%%

  
 %\cite{Jose:1977gm}
\bibitem{Jose:1977gm}
  J.~V.~Jose, L.~P.~Kadanoff, S.~Kirkpatrick, D.~R.~Nelson,
  ``Renormalization, vortices, and symmetry breaking perturbations on the two-dimensional planar model,''
  Phys.\ Rev.\  {\bf B16}, 1217-1241 (1977);
%\cite{Jose:1977gm,135388}

%\bibitem{135388} 
  L.~P.~Kadanoff,
   ``Lattice Coulomb gas representations of two-dimensional problems,''
  J.\ Phys.\ A \ {\bf 11}, 1399  (1978).
  %%CITATION = JPAGB,A11,1399;%%
  

  
  %\cite{Poppitz:2011wy}
\bibitem{Poppitz:2011wy} 
  E.~Poppitz and M.~\" Unsal,
  ``Seiberg-Witten and 'Polyakov-like' magnetic bion confinements are continuously connected,''
  JHEP {\bf 1107}, 082 (2011)
  [arXiv:1105.3969 [hep-th]].
  %%CITATION = ARXIV:1105.3969;%%


%\cite{Dunne:2012zk}

\bibitem{Dunne:2012zk} 
  G.~V.~Dunne and M.~\" Unsal,
   ``Continuity and Resurgence: towards a continuum definition of the $CP^{N-1}$ model,''
  arXiv:1210.3646 [hep-th];
  
  %\cite{Dunne:2012ae}
   G.~V.~Dunne and M.~\" Unsal,
   ``Resurgence and Trans-series in Quantum Field Theory: The $CP^{N-1}$ Model,''
  arXiv:1210.2423 [hep-th].
  %%CITATION = ARXIV:1210.2423;%%
  %%CITATION = ARXIV:1210.3646;%%
  


%\cite{Poppitz:2012sw}
\bibitem{Poppitz:2012sw} 
  E.~Poppitz, T.~Sch\" afer and M.~\" Unsal,
   ``Continuity, Deconfinement, and (Super) Yang-Mills Theory,''
  JHEP {\bf 1210}, 115 (2012)
  [arXiv:1205.0290 [hep-th]].
  %%CITATION = ARXIV:1205.0290;%%
  %%CITATION = HEP-TH/0010201;%%
  %\cite{Anber:2011gn}


\bibitem{Argyres:2012ka} 
  P.~C.~Argyres and M.~\" Unsal,
   ``The semi-classical expansion and resurgence in gauge theories: new perturbative, instanton, bion, and renormalon effects,''
  JHEP {\bf 1208}, 063 (2012)
  [arXiv:1206.1890 [hep-th]].
  %%CITATION = ARXIV:1206.1890;%%
  
  
%\cite{Lee:1997vp}
\bibitem{Lee:1997vp} 
  K.-M.~Lee and P.~Yi,
 ``Monopoles and instantons on partially compactified D-branes,''
  Phys.\ Rev.\ D {\bf 56}, 3711 (1997)
  [hep-th/9702107].
  %%CITATION = HEP-TH/9702107;%%
  %\cite{Argyres:2012ka}
  
%\cite{Anber:2011de}
\bibitem{Anber:2011de} 
  M.~M.~Anber and E.~Poppitz,
   ``Microscopic Structure of Magnetic Bions,''
  JHEP {\bf 1106}, 136 (2011)
  [arXiv:1105.0940 [hep-th]].
  %%CITATION = ARXIV:1105.0940;%%
  
  %\cite{Nelson}
 \bibitem{Nelson} D.R. Nelson, ``Study of melting in two dimensions," Phys. Rev. {\bf B 18}, 2318-2338 (1978).


%\cite{'tHooft:1977hy}
\bibitem{'tHooft:1977hy} 
  G.~'t Hooft,
  ``On the Phase Transition Towards Permanent Quark Confinement,''
  Nucl.\ Phys.\ B {\bf 138}, 1 (1978).
  %%CITATION = NUPHA,B138,1;%%
  
  %\cite{KorthalsAltes:2000gs}
\bibitem{KorthalsAltes:2000gs} 
  C.~Korthals-Altes and A.~Kovner,
  ``Magnetic Z(N) symmetry in hot QCD and the spatial Wilson loop,''
  Phys.\ Rev.\ D {\bf 62}, 096008 (2000)
  [hep-ph/0004052].
  %%CITATION = HEP-PH/0004052;%%
  


%\cite{Smilga:1996cm}\cite{KorthalsAltes:1999xb}
\bibitem{Smilga:1996cm} 
%\cite{Smilga:1993vb}
%\bibitem{Smilga:1993vb} 
  A.~V.~Smilga,
 ``Are Z(N) bubbles really there?,''
  Annals Phys.\  {\bf 234}, 1 (1994);
  %%CITATION = APNYA,234,1;%%


A.~V.~Smilga,
   ``Physics of thermal QCD,''
  Phys.\ Rept.\  {\bf 291}, 1 (1997)
  [hep-ph/9612347];
  %%CITATION = HEP-PH/9612347;%%
  
  %\cite{KorthalsAltes:1999xb}
%\bibitem{KorthalsAltes:1999xb} 
  C.~Korthals-Altes, A.~Kovner and M.~A.~Stephanov,
  ``Spatial 't Hooft loop, hot QCD and Z(N) domain walls,''
  Phys.\ Lett.\ B {\bf 469}, 205 (1999)
  [hep-ph/9909516].
  %%CITATION = HEP-PH/9909516;%%


\bibitem{Nye:2000eg}
  T.~M.~W.~Nye and M.~A.~Singer,
  ``An ${\cal{L}}$$^2$-index theorem for Dirac operators on $\R^3 \times \S^1$,'' J.\ Funct.\ Anal.\ {\bf 177}, 203 (2000); 
  arXiv:math/0009144; 
  
  E.~Poppitz and M.~\" Unsal,
  ``Index theorem for topological excitations on  $\R^3 \times \S^1$ and Chern-Simons
  theory,''
  JHEP {\bf 0903}, 027 (2009)
  [arXiv:0812.2085 [hep-th]].
  %%CITATION = JHEPA,0903,027;%%

\bibitem{Newman-Barkema}
M.~E.~J.~Newman and G.~T.~Barkema,
``Monte Carlo methods in statistical physics,"
Clarendon Press, Oxford (2001), pp. 68-73 and 229-239.
 
%\cite{Tobochnik:1979zz}
\bibitem{Tobochnik:1979zz} 
  J.~Tobochnik and G.~V.~Chester,
``Monte Carlo study of the planar spin model,''
  Phys.\ Rev.\ B {\bf 20}, 3761 (1979).
  %%CITATION = PHRVA,B20,3761;%%


  %\cite{Challa:1986sk}
\bibitem{Challa:1986sk} 
  M.~S.~S.~Challa, D.~P.~Landau and K.~Binder,
 ``Finite size effects at temperature driven first order transitions,''
  Phys.\ Rev.\ B {\bf 34}, 1841 (1986).
  %%CITATION = PHRVA,B34,1841;%%


    
    
   %\cite{Karsch:1998qj}
\bibitem{Karsch:1998qj} 
  F.~Karsch and M.~Lutgemeier,
  ``Deconfinement and chiral symmetry restoration in an SU(3) gauge theory with adjoint fermions,''
  Nucl.\ Phys.\ B {\bf 550}, 449 (1999)
  [hep-lat/9812023].
  %%CITATION = HEP-LAT/9812023;%%
    
 
  \bibitem{Tsvelik}
  A.~M.~Tsvelik, ``Quantum field theory in condensed matter physics," 2nd ed. (Cambridge UP, 2003).


  %%CITATION = HEP-TH/0212230;%%   %\cite{Boyanovsky:1990iw}
\bibitem{Boyanovsky:1990iw}
  D.~Boyanovsky, R.~Holman,
   ``Critical behavior and duality in extended Sine-Gordon theories,''
  Nucl.\ Phys.\  {\bf B358}, 619-653 (1991).
     
%\cite{Lecheminant:2002va}\cite{Lecheminant:2006hj}
\bibitem{Lecheminant:2002va} 
  P.~Lecheminant, A.~O.~Gogolin and A.~A.~Nersesyan,
  ``Criticality in selfdual sine-Gordon models,''
  Nucl.\ Phys.\ B {\bf 639}, 502 (2002)
  [cond-mat/0203294]:
  %%CITATION = COND-MAT/0203294;%%

%\cite{Lecheminant:2006hj}
%\bibitem{Lecheminant:2006hj} 
  P.~Lecheminant,
   ``Nature of the deconfining phase transition in the 2+1-dimensional SU(N) Georgi-Glashow model,''
  Phys.\ Lett.\ B {\bf 648}, 323 (2007)
  [hep-th/0610046].
  %%CITATION = HEP-TH/0610046;%%

  %\cite{Z4paper}
\bibitem{Z4paper}
E.~Rastelli,  S.~Regina, and A.~Tassi
 ``Monte Carlo simulation of a planar rotator model with symmetry-breaking fields", Phys. \ Rev. \ B {\bf 69} 174407, (2004).


 \end{thebibliography}
\end{document}